\documentclass[oneside,12pt]{article}
\usepackage{eurosym}
\usepackage{amsmath}
\usepackage{amsfonts}
\usepackage{graphicx,epsfig}
\usepackage{epstopdf}
\usepackage{amssymb}

\begin{document}

\title{\textbf{\LARGE Cosmic acceleration sourced by modification of gravity
without extra degrees of freedom}}
\author{{\large Abhineet Agarwal}$^{1}${\large , R. Myrzakulov}$^{2}$, 
{\large S. K. J. Pacif}$^{3}$, {\large M. Sami}$^{4}$, {\large Anzhong Wang}$%
^{5}$ \\
$^{1}$\textit{International Institute of Information Technology,}\\
\textit{Gachibowli, Hyderabad,}\\
\textit{Telangana, India}\\
$^{2}$\textit{Eurasian International Center for Theoretical Physics and \ \ }%
\\
\textit{Department of General \& Theoretical Physics,}\\
\textit{Eurasian National University,\ }\\
\textit{Astana 010008, Kazakhstan}\\
$^{3}$\textit{Department of Mathematics,}\\
\textit{School of Advanced Sciences,}\\
\textit{VIT University, Vellore,}\\
\textit{Tamil Nadu 632 014, India}\\
$^{4}$\textit{Centre for Theoretical Physics,}\\
\textit{Jamia Millia Islamia,}\\
\textit{New Delhi 110 025, India}\\
$^{4}$\textit{Maulana Azad National Urdu University,}\\
\textit{Gachibowli, Hyderabad,}\\
\textit{Telangana 500 032, India}\\
$^{4,5}$\textit{Institute for Advanced Physics \& Mathematics,}\\
\textit{Zhejiang University of Technology,\ }\\
\textit{Hangzhou 310032, China}\\
$^{5}$\textit{GCAP-CASPER, Department of Physics,}\\
\textit{Baylor University, Waco }\\
\textit{Texas 76798-7316, USA}\\
\textit{agarwal.abhi93@gmail.com}$^{1}$, \textit{rmyrzakulov@gmail.com}$^{2}$%
,\\
\textit{shibesh.math@gmail.com}$^{3}$, \textit{samijamia@gmail.com}$^{4}$, 
\textit{Anzhong\_Wang@baylor.edu}$^{5}$}

\maketitle

\begin{abstract}
In this paper, we investigate a scenario in which late time cosmic
acceleration might arise due to coupling between dark matter and baryonic
matter without resorting to dark energy or large scale modification of
gravity associated with extra degrees of freedom. The scenario can give rise
to late time acceleration in Jordan frame and no acceleration in Einstein
frame - \textit{generic modification of gravity} caused by disformal
coupling. Using a simple parametrization of the coupling function, in
maximally disformal case, we constrain the model parameters by using the age
constraints due to globular cluster data. We also obtain observational
constraints on the parameters using $H(z)+SNIa+BAO$ datasets. In this case,
we distinguish between phantom and non phantom acceleration and show that
the model can give rise to phantom behavior in a narrow region of parameter
space.
\end{abstract}

\textbf{keywords:} Jordan frame, Einstein frame, disformal coupling, late
time acceleration

\section{Introduction}

It is common belief that late time cosmic acceleration \cite{HZTEAM, SCP,
spergel, seljak} is either sourced by dark energy \cite{DE1, DE2,
DE3,DE4,DE5, DE6,DE7,DE8,DE9,DE10} or by large scale modification of gravity
associated with extra degrees of freedom\footnote{%
we did not mention here large scale modification caused by extra dimensions,
non-local corrections to Einstein-Hilbert action and models based upon f(T)
gravity.}. As for dark energy, it could be represented by cosmological
constant \cite{DE3, carollCC,padmaCC,peeblesCC} or by a slowly rolling
scalar field \cite{quint1, quint2, quint3, quint4, quint5}, both of which
are plagued with the theoretical problem of similar nature. On the other
hand, observations at the background level are quite comfortable with dark
energy scenario. Large scale modification of gravity amounting to Einstein
general theory of relativity along with extra degrees of freedom generically
includes their coupling to matter. The extra degree(s) of freedom, if
massive, should be very light in order to be relevant to late time
acceleration, but would in turn cause a havoc to local physics. Whether the
degrees of freedom are massive or massless, one correspondingly invokes the
chameleon \cite{chamel0, chamel1, chamel2} or Vainstein mechanism \cite%
{vanstein} to suppress them locally. In class of theories, with chameleon
mechanism operative, imposition of local gravity constraints leaves no scope
for self acceleration \cite{Jwang, KBamba}. On the other hand Vainstein
screening sets in dynamically through non-linear derivative interaction and
sounds superior to chameleon mechanism. For instance, it is at the heart of
massive theories of gravity. Furthermore,
In the case of massive gravity \textit{a la} dRGT \cite{dRGT1, dRGT2}, the
scalar belongs to Galileon type \cite{galileon} which gets screened via
Vainstein screening. Unfortunately, FRW cosmology is absent in this theory.
Promoting the latter to bi-gravity, might address the problem but then
issues related to Higuchi bound pope in. It would be fare to say that, at
present, a consistent model of large scale modification of gravity,
associated with extra degrees of freedom, relevant to late time
acceleration, is not known.In these scenarios, in the decoupling limit
relevant to local physics, vector degrees of freedom decouple whereas the
longitudinal (scalar) one couples to matter with the universal coupling.

In
a recent review \cite{EXTENDED}, a model-independent approach was considered
to tackle the dark energy/modified gravity problem wherein $f(R)$ and $f(T)$
theories were explored in different formalisms and the role of conformal
transformations in the Einstein and Jordan frames was clarified, see also Ref.\cite{CAPOZZI} on related issues. 

Clearly, it should be interesting to look for the third alternative leaving
aside the exotic matter or extra degrees of freedom responsible for large
scale modifications of gravity. Recently, a novel mechanism was proposed to
do the needful, namely, interaction between baryonic and dark matter was
shown to give rise to late time acceleration \cite{KHOURY2016}. In this
case, as demonstrated in Ref. \cite{KHOURY2016}, the purely conformal
coupling is disfavored by the stability criteria. However, the maximally
disformal coupling can give rise to late time cosmic acceleration in Jordan
frame and no acceleration in the Einstein frame as the energy density of
both matter components taken together follows standard conservation and
redshifts as usual. On the other hand, in the Jordan frame, the matter
components are not coupled but dynamics is modified such that late time
acceleration might arise in this frame. In our opinion, this is a remarkable
possibility of generic modification of gravity$-$ acceleration in Jordan
frame and no acceleration in Einstein frame.\newline
In this paper, we further investigate the proposal of Ref. \cite{KHOURY2016}%
. We use simple parametrization of metric function in a maximally disformal
case and constrain the model parameters using the age constraints due to
globular cluster data. We also find constraints on the parameters using $%
H(z)+SNIa+BAO$ datasets for phantom and non phantom acceleration.

\section{Interaction between dark matter and baryons}

In this section, we briefly revisit the scenario introduced in Ref. \cite%
{KHOURY2016}. As pointed out in the introduction, we shall consider a
general coupling, to be specified later, between baryonic matter and dark
matter. To this effect, we shall use the following action in the Einstein
frame Ref. \cite{KHOURY2016}, 
\begin{equation}
\mathcal{L}=\frac{1}{16\pi G}\sqrt{-g}\mathcal{R}+\mathcal{L}_{DM}[g_{\mu
\nu }]+\mathcal{L}_{b}[\tilde{g}_{\mu \nu }]\text{,}  \label{2}
\end{equation}%
where $\mathcal{L}_{DM}$ describes the dark matter which is minimally
coupled; $\mathcal{L}_{b}$ the Lagrangian for baryonic matter that couples
to dark matter through Jordan frame metric $\tilde{g}_{\mu \nu }$ which is
constructed from the Einstein frame metric $g_{\mu \nu }$ and parameters
that characterize the dark matter. From here onwards, quantities with a
overhead tilde would be associated with Jordan frame.

In order to fix the Lagrangian for dark matter, for simplicity, we assume
dark matter to be a perfect fluid which is legitimate on linear scales we
are interested in. In this case, dark matter can be described by the
Lagrangian of a single scalar field, 
\begin{equation}
\mathcal{L}_{DM}=\sqrt{-g}P(X),  \label{Ldm}
\end{equation}%
where $X=-g^{\mu \nu }\partial _{\mu }\Theta \partial _{\nu }\Theta $, ($%
\Theta $ being the dark matter field). For the action (\ref{Ldm}), the
stress tensor is given by 
\begin{equation}
T_{\mu \nu }=2P,_{X}\partial _{\mu }\Theta \partial _{\nu }\Theta +Pg_{\mu
\nu }\text{,}  \label{2a}
\end{equation}%
which has the form of energy momentum tensor of perfect fluid $T_{\mu \nu
}=(\rho _{DM}+P_{DM})u_{\mu }u_{\nu }+P_{DM}g_{\mu \nu }$, provided that we
make the following identification for the density and pressure of dark
matter, 
\begin{equation}
\rho _{DM}=2P,_{X}(X)X-P(X)\text{ , }P_{DM}=P(X)\text{ \ , \ }u_{\mu }=-%
\frac{1}{\sqrt{X}}\partial _{\mu }\Theta \text{.}  \label{3}
\end{equation}%
As mentioned before, the baryonic matter couples to dark matter via the
Jordan frame metric $\tilde{g}_{\mu \nu }$ which is constructed from the
Einstein frame metric $g_{\mu \nu }$ and components of dark matter. With the
assumption that dark matter is perfect fluid, the most general form of $%
\tilde{g}_{\mu \nu }$ is given by

\begin{eqnarray}
&&\tilde{g}_{\mu \nu }=-Q^{2}(X)u_{\mu }u_{\nu }+R^{2}(X)(g_{\mu \nu
}+u_{\mu }u_{\nu })\text{,}  \label{4} \\
&&\tilde{g}_{\mu \nu }=R^{2}(X)g_{\mu \nu }+S(X)\partial _{\mu }\Theta
\partial _{\nu }\Theta ;~~S(X)\equiv \frac{R^{2}(X)-Q^{2}(X)}{X}\text{,}
\label{4A}
\end{eqnarray}%
where $R$ and $Q$ are arbitrary functions to begin with and determinants of
both the Einstein frame and Jordan frame metrics are related by $\sqrt{- 
\tilde{g}}=QR^{3}\sqrt{-g}$. With the above specialization of matter
components, the equations of motion for DM can be obtained by varying the
action with respect to the dark matter field $\Theta $, Ref. \cite%
{KHOURY2016},

\begin{equation}
\partial _{\nu }\left( \left[ 2P,_{X}+QR^{3}\tilde{T}_{b}^{\alpha \beta
}(2RR,_{X}g_{\alpha \beta }+S,_{X}\partial _{\alpha }\Theta \partial _{\beta
}\Theta )g^{\mu \nu }-QR^{3}S\tilde{T}_{b}^{\mu \nu }\right] \sqrt{-g}%
\partial _{\mu }\Theta \right) =0\text{,}  \label{5}
\end{equation}%
where $\tilde{T}_{b}^{\mu \nu }$ is the Jordan frame energy-momentum tensor
for baryons given by,

\begin{equation}
\tilde{T}_{b}^{\mu \nu }=\frac{2}{\sqrt{-\tilde{g}}}\frac{\delta \mathcal{L}
_{b}}{\delta \tilde{g}_{\mu \nu }}\text{.}  \label{6}
\end{equation}%
In Jordan frame, the matter components are not coupled though the dynamics
might look complicated, see Appendix-1 for details. Thus both the components
are separately conserved. In particular, 
\begin{equation}
\tilde{\triangledown}_{\mu }\tilde{T}_{b}^{\mu \nu }=0\text{.}  \label{6a}
\end{equation}%
However, in the Einstein frame, dark matter and baryonic matter do not
conserve separately due to the presence of coupling. Nevertheless, the total
energy density of both the components taken together should still follow
standard conservation law. To this effect, following Ref. \cite{KHOURY2016},
let us first write down the Einstein equation, varying the action with
respect to $g_{\mu \nu }$, we obtain

\begin{equation}
G_{\mu \nu }=8\pi G\left[ T_{\mu \nu }+QR^{3}\tilde{T}_{b}^{k\lambda }\left(
R^{2}g_{k\mu }g_{\lambda \nu }+(2RR,_{X}g_{k\lambda }+S,_{X}\partial
_{k}\Theta \partial _{\lambda }\Theta )\partial _{\mu }\Theta \partial _{\nu
}\Theta \right) \right] \text{,}  \label{7}
\end{equation}%
where the DM stress-energy tensor $T_{\mu \nu }$ is given in (\ref{2a}). The
second term on LHS in the above equation is the effective energy momentum
tensor in the Einstein frame.

\section{Cosmological dynamics in presence of coupling}

The scenario under consideration is based upon the assumption of coupling
between baryonic and dark matter components, introduced phenomenologically,
in the Einstein frame. In order to investigate the observational
consequences of (\ref{2}), we shall, hereafter, specialize to homogeneous
and isotropic Universe on a spatially-flat FRW (Friedmann-Robertson-Walker)
background

\begin{equation}
ds^{2}=-dt^{2}+a^{2}(t)\left( dx^{2}+dy^{2}+dz^{2}\right) \text{.}  \label{8}
\end{equation}%
Since the metric functions $R$ and $Q$ are functions of the scale factor
only on the FRW background, the Jordan frame metric tensor then takes the
form,

\begin{equation}
\tilde{g}_{\mu \nu }=diag\left(
-Q^{2}(a),R^{2}(a)a^{2},R^{2}(a)a^{2},R^{2}(a)a^{2}\right) \text{.}
\label{9}
\end{equation}%
The arbitrary functions $Q(a)$ and $R(a)$ need to be specified such that the
thermal history, known to good accuracy, is left intact. Thus, in the early
Universe, gravity should become standard so that in the limit of high matter
density, $Q,$ and $R$ become constant which without the loss of generality
could be taken unity. This condition has to be imposed phenomenologically
which also has implications for local physics. Indeed, the said choice would
reduce the model to Einstein gravity in high density regime adhering to
local gravity constraints. Secondly, at late times $Q$ and $R$ should be
chosen such that the baryons experience accelerated expansion. Let us note
that, since total energy density of matter in the Einstein frame follow
standard conservation, such a scheme can not give rise to acceleration in
this frame. We can expect acceleration in the Jordan frame only provided
that we make appropriate choices. Indeed, the Jordan frame scale factor
dubbed \textit{physical scale factor} and its counter part in the
Einstein-frame are related as

\begin{equation}
\tilde{a}=Ra\text{.}  \label{10}
\end{equation}%
Since, $Q,$ $R\rightarrow 1$ in the early Universe, $\tilde{a}\rightarrow a$
at early times. Although the Einstein frame scale factor is always
decelerating, the expansion in Jordan frame governed by $\tilde{a}$ might
exhibit acceleration if $R$ grows sufficiently fast at late times. In
generic case, $R$ is concave up beginning from $R=1$ at early times. In that
case fast growth of $R$ might compensate the effect of deceleration of $a(t)$
making $\ddot{\tilde{a}}$ positive. Indeed, 
\begin{equation}
\ddot{\tilde{a}}=\ddot{R}a+2\dot{R}\dot{a}+R\ddot{a}\text{.}  \label{aceq}
\end{equation}%
The last term in Eq. (\ref{aceq}) is negative and in case $R$ is concave up,
the first term is positive. Hence, if $R$ grows fast at late times, the
second term can compensate the effect of deceleration in Einstein frame
giving rise to acceleration in the Jordan frame. In this case, acceleration
is completely removed by disformal transformation which signifies that
acceleration is generic feature of large scale modification of gravity in
the Jordan frame.

Assuming the baryon component to be a perfect fluid, 
\begin{equation}
\tilde{T}_{b}^{\mu \nu }=(\tilde{\rho}_{b}+\tilde{P}_{b})\tilde{u}_{b}^{\mu }%
\tilde{u}_{b}^{\nu }+\tilde{P}_{b}\tilde{g}^{\mu \nu }~;~~~\tilde{g}^{\mu
\nu }\tilde{u}_{b}^{\mu }\tilde{u}_{b}^{\nu }=-1\text{.}  \label{12}
\end{equation}%
Eq. (\ref{6a}) then gives the standard continuity equation

\begin{equation}
\frac{d\tilde{\rho}_{b}}{d\ln \tilde{a}}=-3(\tilde{\rho}_{b}+\tilde{P}_{b})%
\text{.}  \label{13}
\end{equation}

The dark matter equation (\ref{5}) reduces to \cite{KHOURY2016}, 
\begin{equation}
\frac{d}{dt}\left( \left[ -P,_{X}+QR^{3}\left( \frac{Q,_{X}}{Q}\tilde{\rho}%
_{b}-3\frac{R,_{X}}{R}\tilde{P}_{b}\right) \right] a^{3}\dot{\Theta}\right)
=0\text{,}  \label{14}
\end{equation}%
where we have used $g_{k\lambda }\tilde{T}_{b}^{k\lambda }=-Q^{-2}\tilde{\rho%
}_{b}+3R^{-2}\tilde{P}_{b}$ and assumed, $\dot{\Theta}>0$. Using equation (%
\ref{3}) and integrating (\ref{14}), we have

\begin{equation}
\rho _{DM}=\Lambda _{DM}^{4}\sqrt{\frac{X}{X_{eq}}}\left( \frac{a_{eq}}{a}%
\right) ^{3}-P+2XQR^{3}\left( \frac{Q,_{X}}{Q}\tilde{\rho}_{b}-3\frac{R,_{X}%
}{R}\tilde{P}_{b}\right) \text{,}  \label{15}
\end{equation}%
where the subscript `$eq$' indicates matter-radiation equality. $\Lambda
_{DM}^{4}$ is identified as the DM mass density at radiation-matter
equality. The Friedmann equation can be derived from equation (\ref{7}) as

\begin{equation}
3H^{2}=8\pi G\left( \rho _{DM}+\rho _{b}\right) \text{,}  \label{16}
\end{equation}%
where an effective Einstein-frame baryon density is defined as

\begin{equation}
\rho _{b}=QR^{3}\left( \tilde{\rho}_{b}\left( 1-2X\frac{Q,_{X}}{Q}\right) +6X%
\frac{R,_{X}}{R}\tilde{P}_{b}\right) \text{.}  \label{17}
\end{equation}

Using Eqs. (\ref{15}) and (\ref{17}) in Eq. (\ref{16}), the Friedmann
equation becomes

\begin{equation}
3H^{2}=8\pi G\left( \Lambda _{DM}^{4}\sqrt{\frac{X}{X_{eq}}}\left( \frac{
a_{eq}}{a}\right) ^{3}-P+QR^{3}\tilde{\rho}_{b}\right) \text{.}  \label{18}
\end{equation}

Radiation do not couple to either baryon or DM. The acceleration equation
can be derived as

\begin{equation}
2\frac{\ddot{a}}{a}+H^{2}=-8\pi G(P+P_{b})\text{,}  \label{19}
\end{equation}%
where the effective baryon pressure is

\begin{equation}
P_{b}\equiv QR^{3}\tilde{P}_{b}\text{.}  \label{20}
\end{equation}

Now, we have three field equations (\ref{14}), (\ref{18}) and (\ref{19}) out
of which two are independent.

Hereafter, we shall assume matter to be pressureless, $\tilde{P}_{b}\simeq 0%
\text{ and }P\ll 2XP,_{X}\label{21}$. Since both the matter components in
Jordan frame conserve separately, they follow the standard conservation.
Indeed, using Eqs. (\ref{20}) in Eq. (\ref{13}), one obtains, $\tilde{\rho}%
_{b}\sim \tilde{a}^{-3}$ as it should be in absence of pressure. Baryon
matter density can conveniently be written as, 
\begin{equation}
\tilde{\rho}_{b}=\frac{\Lambda _{b}^{4}}{R^{3}}\left( \frac{a_{eq}}{a}%
\right) ^{3}\text{.}  \label{21a}
\end{equation}%
\qquad

Since the function $R=1$ in the early Universe, $\Lambda _{b}^{4}$ can be
treated as the baryon mass density at equality. For zero pressure, Eq. (\ref%
{19}) leads to $a(t)\sim t^{\frac{2}{3}}$ and the background is identical to
matter dominated which is valid from mater-radiation equality up to the
present time irrespective of the dark matter-baryon coupling. The total
energy density can be obtained from Eq. (\ref{18}) after substituting Eq. (%
\ref{21a}) as

\begin{equation}
\rho _{Total}\equiv \frac{3H^{2}}{8\pi G}\simeq \left[ \Lambda _{DM}^{4}%
\sqrt{\frac{X}{X_{eq}}}+\Lambda _{b}^{4}\right] \left( \frac{a_{eq}}{a}%
\right) ^{3}\text{,}  \label{22}
\end{equation}%
The total energy density in absence of pressure in the Einstein frame should
follow the standard conservation implying that $\rho _{Total}\sim a^{-3}$.
It is important to note that the conservation holds for an arbitrary $Q$
which means that the term within square bracket in Eq. (\ref{22}) is time
independent for any $Q(X)$. Clearly, coupling does not give rise to
acceleration in Einstein-frame. However, in the Jordan frame, where the
matter components are not directly coupled but gravity is modified, we might
achieve acceleration at late times. We do not take this path, we shall
rather directly work with the Jordan frame metric with a suitable
parametrization\footnote{%
In the generic case, $Q$ can be fixed to unity.} of metric function $R$.

Let us define the density parameters as

\begin{equation}
\Omega _{DM}=\frac{\rho _{DM}}{\frac{3H^{2}}{8\pi G}}\text{, \ \ }\Omega
_{b}=\frac{\rho _{b}}{\frac{3H^{2}}{8\pi G}}\text{.}  \label{24a}
\end{equation}

In case of zero pressure, we have,

\begin{equation}
\Omega _{DM}\equiv \Omega _{DM}^{(0)}\left( \frac{a_{0}}{a}\right)
^{3}\left( \frac{H_{0}}{H}\right) ^{2}\text{, \ \ }\Omega _{b}\equiv \Omega
_{b}^{(0)}\left( \frac{a_{0}}{a}\right) ^{3}\left( \frac{H_{0}}{H}\right)
^{2}\text{.}  \label{24B}
\end{equation}

The Friedmann equation, 
\begin{equation}
H^{2}=H_{0}^{2}\left[ \Omega _{DM}^{(0)}(1+z)^{3}+\Omega _{b}^{(0)}(1+z)^{3}%
\right] \text{,}  \label{24c}
\end{equation}%
has the standard equation of flat FRW cosmology with cold matter(dark
matter+baryonic matter); subscript `$0$' indicates the value of the quantity
at present epoch as usual. We denote the present time mass densities as
superscript `$(0)$' for respective quantity.

As mentioned before, the two coupling functions $Q$ and $R$ can be chosen
suitably such that $Q,$ $R\rightarrow 1$ in the early Universe and both $a$
and $\tilde{a}$ experience deceleration where as at late times the coupling
function $R$ grows sufficiently large such that the physical scale factor $%
\tilde{a}$ \ experience acceleration. Let us distinguishes two choices of $Q$
and $R$. The conformal coupling that corresponds to $Q(X)=R(X)$ and the
disformal coupling $Q(X)\neq R(X)$. The stability condition for $%
0<c_{s}^{2}\leq 1$ requires fine tuning in case of conformal coupling and
sounds unnatural. Indeed, the stability condition, $\rho _{DM}>>Q^{4}\rho
_{b}/c_{DM}^{2}$ requires unnatural choice for $Q$ corresponding to a given
value of $c_{DM}$, see Ref. \cite{KHOURY2016} for details. The stability
condition holds for the disformal coupling in case of $Q=1$ dubbed maximally
disformal . In this case, one is left with one function $R$ which can be
conveniently parametrized and compared with data. We will focus on a simple
parametrization of this function involving two parameters that can be
constrained from data.

\subsection{Simple parametrization and cosmological dynamics in Jordan frame}

It is clear from Eq. (\ref{24c}) that the scale factor in Einstein frame
does not experience acceleration. We should look for such a possibility in
Jordan frame where matter components are not coupled but dynamics is
modified. In order to progress further, we need to specify $R(a)$.
Equivalently, we can also parametrize $a$ in terms of physical scale factor $%
\tilde{a}$. In the discussion to follow, we consider the following simple
form of $a(\tilde{a})$, 
\begin{equation}
a(\tilde{a})=\tilde{a}+\alpha \tilde{a}^{2}+\beta \tilde{a}^{3}\text{,}
\label{25}
\end{equation}%
where $\alpha $ and $\beta $ are two parameters of the model to be
determined from observational data. The functional form is chosen in such a
way that in the early times $a=\tilde{a}$ such that the distinction between
the two frames disappear at early times. We also note that this simple
functional form, as series expansion in $a=\tilde{a}$ is valid up to the
present time, $\tilde{a}\leq 1$. Extrapolating (\ref{25}) to future
evolution might bring in undesirable features. For simplicity, we shall
restrict our discussion to the polynomial of third degree in $\tilde{a}$ in (%
\ref{25}) which is convenient for comparing results with observation. The
first and second derivatives of the scale factor are given by 
\begin{equation}
\dot{a}=(1+2\alpha \tilde{a}+3\beta \tilde{a}^{2})\dot{\tilde{a}}\text{ \ ,
\ }\ddot{a}=(1+2\alpha \tilde{a}+3\beta \tilde{a}^{2})\ddot{\tilde{a}}%
+(2\alpha +6\beta \tilde{a})\dot{\tilde{a}}^{2}\text{,}  \label{26}
\end{equation}%
where $\dot{\tilde{a}}=\frac{d\tilde{a}}{dt}$. Using these,\textbf{\ } we
can obtain the Hubble parameter and deceleration parameter in Jordan-frame as

\begin{equation}
\tilde{H}=\frac{\dot{\tilde{a}}}{\tilde{a}}=\left( \frac{1+\alpha \tilde{a}
+\beta \tilde{a}^{2}}{1+2\alpha \tilde{a}+3\beta \tilde{a}^{2}}\right) H%
\text{,}  \label{27}
\end{equation}

and%
\begin{equation}
\tilde{q}=-\frac{\tilde{a}\ddot{\tilde{a}}}{\dot{\tilde{a}}^{2}}=\frac{1}{2}%
\left( \frac{1+2\alpha \tilde{a}+3\beta \tilde{a}^{2}}{1+\alpha \tilde{a}
+\beta \tilde{a}^{2}}\right) +\left( \frac{2\tilde{a}(\alpha +3\beta \tilde{%
a })}{1+2\alpha \tilde{a}+3\beta \tilde{a}^{2}}\right) \text{.}  \label{28}
\end{equation}

($\because $ In Einstein frame, for zero pressure Eq. (\ref{19}) gives $q=%
\frac{1}{2}$). Also, using equation (\ref{26}) along with the field
equations, we obtain

\begin{equation}
2\left( \frac{1+2\alpha \tilde{a}+3\beta \tilde{a}^{2}}{1+\alpha \tilde{a}
+\beta \tilde{a}^{2}}\right) \frac{\ddot{\tilde{a}}}{\tilde{a}}+\frac{
2(\alpha +3\beta \tilde{a})}{1+\alpha \tilde{a}+\beta \tilde{a}^{2}}\frac{ 
\dot{\tilde{a}}^{2}}{\tilde{a}^{2}}=-\frac{8\pi G}{3}\left[ \Lambda _{DM}^{4}%
\sqrt{\frac{X}{X_{eq}}}+\Lambda _{b}^{4}\right] \left( \frac{a_{eq}}{a}%
\right) ^{3}\text{,}  \label{29}
\end{equation}%
which gives an expression of deceleration parameter in Jordan frame as

\begin{equation}
\tilde{q}=\frac{\Omega _{Total}}{2}\left( \frac{1+2\alpha \tilde{a}+3\beta 
\tilde{a}^{2}}{1+\alpha \tilde{a}+\beta \tilde{a}^{2}}\right) +\left( \frac{%
2 \tilde{a}(\alpha +3\beta \tilde{a})}{1+2\alpha \tilde{a}+3\beta \tilde{a}%
^{2}}\right) \text{,}  \label{30}
\end{equation}%
where we have used the equation (\ref{22}). Equations (\ref{28}) and (\ref%
{30}) are same because $\Omega _{Total}=1$.

The Hubble parameter (Eq. (\ref{27})) can be represented as%
\begin{equation}
\frac{\tilde{H}}{\tilde{H}_{0}}=\left( \frac{1+2\alpha +3\beta }{1+\alpha
+\beta }\right) \left( \frac{1+\alpha \tilde{a}+\beta \tilde{a}^{2}}{
1+2\alpha \tilde{a}+3\beta \tilde{a}^{2}}\right) \left( \frac{H}{H_{0}}%
\right) \text{.}  \label{32a}
\end{equation}

Since, for pressureless matter, $H(a)=H_{0}\left( \frac{a_{0}}{a}\right) ^{ 
\frac{3}{2}}$ and $a_{0}=(1+\alpha +\beta )$, using Eq. (\ref{25}), Eq. (\ref%
{32a}) can be also be cast as%
\begin{equation}
\tilde{H}(\tilde{a})=\tilde{H}_{0}\frac{(1+\alpha +\beta )^{\frac{1}{2}
}(1+2\alpha +3\beta )}{\tilde{a}^{\frac{3}{2}}\left[ 1+\alpha \tilde{a}
+\beta \tilde{a}^{2}\right] ^{\frac{1}{2}}\left[ 1+2\alpha \tilde{a}+3\beta 
\tilde{a}^{2}\right] }\text{.}  \label{32b}
\end{equation}

From Eg. (\ref{10}), we have, 
\begin{equation}
\frac{a}{\tilde{a}}=\frac{1}{R}\text{.}  \label{23}
\end{equation}

We set the physical scale factor at present to be $1$ i.e. $\tilde{a}_{0}=1$%
, but this implies that $a_{0}\neq 1$. The redshifts in both the Einstein
frame and Jordan frame are then related as

\begin{equation}
\tilde{a}=\frac{\tilde{a}_{0}}{1+\tilde{z}}\text{ , \ \ }a=\frac{a_{0}}{1+z}%
\text{.}  \label{24}
\end{equation}%
The present time corresponds to $\tilde{z}=z=0$.

Now, using equations (\ref{24}) and (\ref{25}), the Hubble parameter can be
expressed in terms of redshift $\tilde{z}$ in Jordan frame as 
\begin{equation}
\tilde{H}(\tilde{z})=\tilde{H}_{0}\frac{(1+\alpha +\beta )^{\frac{1}{2}
}(1+2\alpha +3\beta )\left( 1+\tilde{z}\right) ^{\frac{9}{2}}}{\left[ \left(
1+\tilde{z}\right) ^{2}+\alpha \left( 1+\tilde{z}\right) +\beta \right] ^{ 
\frac{1}{2}}\left[ \left( 1+\tilde{z}\right) ^{2}+2\alpha \left( 1+\tilde{z}
\right) +3\beta \right] }\text{.}  \label{33}
\end{equation}

\begin{figure}[tbph]
\centering
\includegraphics[width=0.50\textwidth]{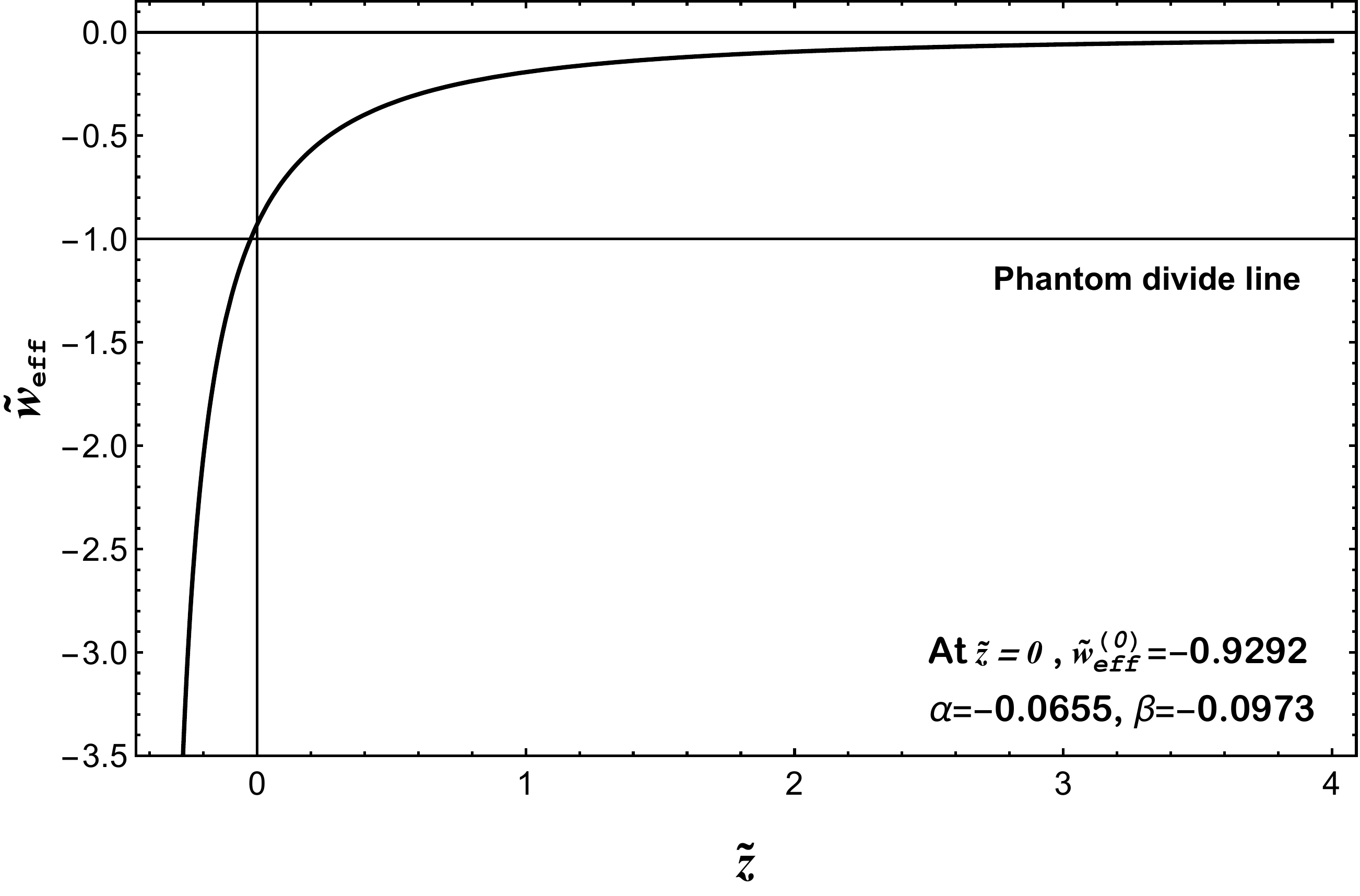}
\caption{Figure shows the equation of state parameter versus the redshift in
Jordan frame. Parameters are set such that we have non-phantom behavior at
present; slight adjustment of parameters can give rise to phantom behavior
consistent with the observation. Even if the present behavior is set to be
non-phantom, equation of state evolves to large negative values in future as 
$\tilde{z}\rightarrow \tilde{z_{s}}=-0.3902$. Such a behavior is a clear
manifestation of a future singularity which is specific to (\protect\ref{25}
).}
\label{fig:weff}
\end{figure}

Let us also quote the expression for the equation of state parameter versus
the redshift ($\tilde{z}$) in Jordan frame as, 
\begin{equation}
\tilde{w}_{eff}(\tilde{z})=\frac{\alpha \left( 5+6\alpha +5\tilde{z}\right)
(1+\tilde{z})^{2}+\beta (14+23\alpha +14\tilde{z})(1+\tilde{z})+18\beta ^{2}%
}{3\{(1+\tilde{z})^{2}+\alpha (1+\tilde{z})+\beta \}\{(1+\tilde{z}
)^{2}+2\alpha (1+\tilde{z})+3\beta \}}\text{,}  \label{w}
\end{equation}%
which clearly mimics cold matter in the limit of large redshift. In the
scheme of two parameters, we have derived the effective equation of state
induced by coupling between two known components of matter. Analytical
expression (\ref{w}) can directly be used for finding observational
constraints on the model. Let us note that for generic negative values of
parameters, the denominator in (\ref{w}) vanishes for certain negative
values of the redshift around which the numerator is always negative. Hence,
if (\ref{25}) is extrapolated to near future, the equation of state
parameter after assuming the observed value at the present epoch, would
attain larger and larger negative values. It is clear that the underlying
system would then always evolve to phantom in future (even if we set the
non-phantom behavior at present, see Fig. \ref{fig:weff}). This is, however,
not the generic feature of the scenario but rather the artifact of
parametrization (\ref{25}).

Expressions (\ref{33}) and (\ref{w}) are the important results of our
analyses to be used for further manipulations. At the onset (\ref{33}) (for
non-vanishing couplings $\alpha $ and $\beta $ which imbibe new physics)
does not look like the expression for Hubble parameter with standard matter
plus an exotic (dark) fluid. Since, suitable, negative values of the
couplings can give rise to late time acceleration, it should be possible to
extract the required information from (\ref{33}) by putting it in a
convenient form. However, we shall first test it for the resolution of age
problem in the hot big bang model.

\subsection{Age of the Universe}

It is well known that the FRW cosmology is plagued with age crises if
Universe is inhabited by the standard form of matter alone which contributes
to deceleration as gravity is attractive. Secondly, more than half the
contribution to age of Universe comes from late stages, namely, evolution
from $z=0$ to $z=1$. If hypothetically, we ignore gravity, then Hubble law
gives rise to $t_{0}=1/H_{0}$, thereby the disturbing factor of $2/3$ in the
formula for the age of Universe is contributed by gravity in presence of
normal matter. Since we can not do away with normal matter (cold dark
matter+baryonic matter to be called as cold matter or standard matter in the
further discussion)\footnote{%
Radiation is also included in the standard matter if relevant, radiation,
however, does not contribute to age of Universe.}, the only way out in the
standard frame work is provided by the assumption of existence of an exotic
form of matter, repulsive in nature, that dominates the late Universe and
can compensate deceleration caused by the standard matter for which gravity
is attractive. Clearly, the latter would slow down the expansion rate at
late times improving the age of Universe. Since in the present scenario, we
have only standard matter, it is desirable to check for the age of Universe.
Using expression (\ref{32b}), we can compute the age of Universe in the
model under consideration in Jordan frame\footnote{%
We should emphasize that the age of Universe computed in the Einstein frame
would still be, $t_{0}=2/3H_{0}$ as there is no acceleration in this case.},

\begin{equation}
\tilde{t}_{0}=\frac{1}{\tilde{H}_{0}}\int\limits_{0}^{\infty }\frac{\left[
\left( 1+\tilde{z}\right) ^{2}+\alpha \left( 1+\tilde{z}\right) +\beta %
\right] ^{\frac{1}{2}}\left[ \left( 1+\tilde{z}\right) ^{2}+2\alpha \left(
1+ \tilde{z}\right) +3\beta \right] }{(1+\alpha +\beta )^{\frac{1}{2}
}(1+2\alpha +3\beta )\left( 1+\tilde{z}\right) ^{\frac{11}{2}}}d\tilde{z}%
\text{.}  \label{33A}
\end{equation}

\begin{figure}[h]
\centering
\includegraphics[scale=.55]{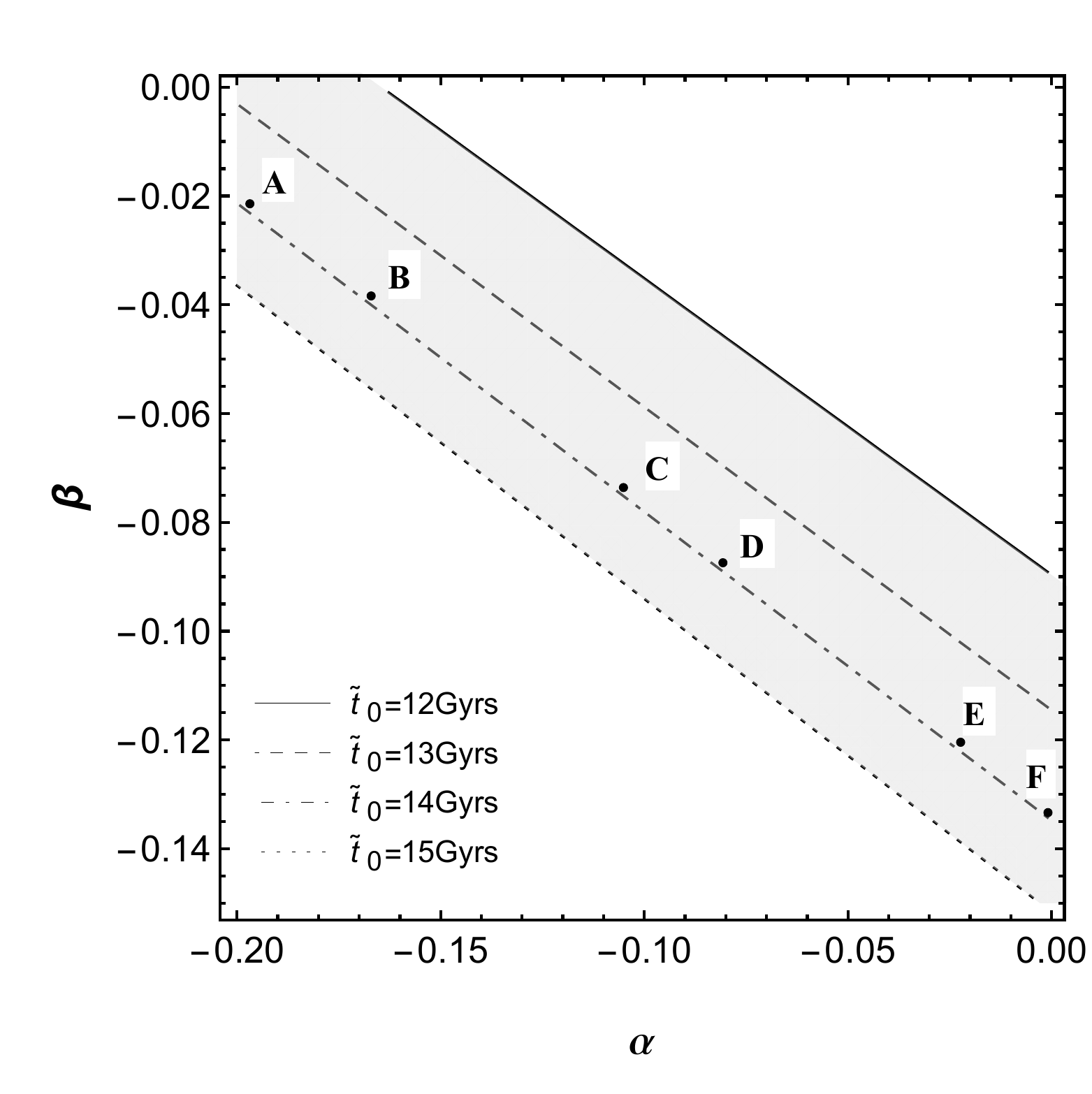}
\caption{Different values of age of the Universe have been plotted in the $%
\protect\alpha \protect\beta -$plane. The solid line, dashed, dot-dashed and
dotted lines correspond to $\tilde{t}_{0}(\tilde{H}_{0}/67.8)=12$ Gyrs, $13$
Gyrs, $14$ Gyrs and $15$ Gyrs, respectively. The shaded region shows the
allowed values of $\protect\alpha $ and $\protect\beta $ for the
aforementioned range of age of the Universe.}
\label{fig:Age}
\end{figure}

In absence of coupling, $\alpha =\beta =0$, we have $\tilde{t}_{0}=t_{0}={2}/%
{3H_{0}}$ as should be. The essence of acceleration lies in the
non-vanishing values of coupling parameters $\alpha $ \& $\beta $. In Fig. %
\ref{fig:Age}, we show parameter range consistent with globular cluster data
on the age of Universe. The range is found using the expression (\ref{33A})
with $\tilde{H}_{0}=67.8$ $km/s/Mpc$. We have plotted age of the Universe
ranging from $12$ Gyrs to $15$ Gyrs in the $\alpha \beta -$plane. The shaded
region shows the allowed values of $\alpha $ \& $\beta $ corresponding to
the range of age of the Universe as dictated by globular cluster data. The
parameters $\alpha $ \& $\beta $ can be set to get higher age of the
Universe with large acceleration.

\qquad

In order to understand the underlying physics, let us focus on the
dot-dashed line corresponding to $t_{0}=14$ Gyrs. If we move down around
this line towards smaller values of $\alpha $ (large values of $\beta $)
(numerically), we observe an interesting pattern, namely, $\Omega
_{Meff}^{(0)}$ increases whereas $\tilde{w}_{eff}^{(0)}$ decreases towards
larger negative values, see Table \ref{Age-tab}. Larger values of $\Omega
_{Meff}^{(0)}$ amounts to more deceleration which decreases the age of the
Universe. The effect is compensated by stronger repulsive effect caused by
larger negative values of the effective equation of state such that the age
of the Universe does not change. This pattern clearly shows the underlying
importance of late time cosmic acceleration for the resolution of age crisis
in the standard model of cosmology. We should, however, admit that age
considerations give broad constraints on the parameters. One certainly
requires other data sets for better constraints.\newline

\begin{table}[tbp]
\begin{center}
\begin{tabular}{|c|c|c|c|}
\hline
$\left( \alpha ,\beta \right) $ & $w_{eff}^{(0)}$ & $\Omega _{Meff}^{(0)}$ & 
$\Omega _{X}^{(0)}$ \\ \hline
\multicolumn{1}{|l|}{$A(-0.1952,-0.0225)$} & $-0.7484$ & $0.2298$ & $0.7701$
\\ \hline
\multicolumn{1}{|l|}{$B(-0.1633,-0.0407)$} & $-0.7927$ & $0.2419$ & $0.7581$
\\ \hline
\multicolumn{1}{|l|}{$C(-0.1075,-0.0720)$} & $-0.8602$ & $0.2656$ & $0.7344$
\\ \hline
\multicolumn{1}{|l|}{$D(-0.0814,-0.0864)$} & $-0.8875$ & $0.2780$ & $0.7220$
\\ \hline
\multicolumn{1}{|l|}{$E(-0.0340,-0.1140)$} & $-0.9522$ & $0.2965$ & $0.7035$
\\ \hline
\multicolumn{1}{|l|}{$F(-0.0024,-0.1343)$} & $-1.0170$ & $0.3028$ & $0.6972$
\\ \hline
\end{tabular}%
\end{center}
\caption{Table lists points $A\rightarrow F$ on $\protect\alpha \protect%
\beta -$plane around $t_{0}=14$ Gyr line from top to bottom with the
corresponding values of equation of state and density parameters. Table
shows that weakening deceleration corresponds to weakening of acceleration.}
\label{Age-tab}
\end{table}

\subsection{Two fluid representation and connection to the standard lore}

Since coupling between non components of matter in Einstein frame gives rise
to acceleration in Jordan frame, it might be possible to define an effective
hypothetical fluid that would mimic \textit{dark energy}. Indeed, equation (%
\ref{33}) can also be cast as\footnote{%
We have put the square of rational expression (\ref{33}) in a convenient
form isolating the term proportional to $(1+\tilde{z})^{3}$ that can be
identified with the effective fractional density of cold matter and the rest
is pushed to a hypothetical matter expected to mimic dark energy.}.

\begin{equation}
\frac{\tilde{H}^{2}}{\tilde{H}_{0}^{2}}=A(\alpha ,\beta )(1+\tilde{z}%
)^{3}+A(\alpha ,\beta )f(\tilde{z})\text{,}  \label{34}
\end{equation}%
where $\left( \Omega _{DM}^{(0)}+\Omega _{b}^{(0)}\right) =1$, in the case
under consideration in Einstein frame and functions $A$ and $f$ are given
by, 
\begin{eqnarray}
A(\alpha ,\beta ) &=&(1+\alpha +\beta )(1+2\alpha +3\beta )^{2}\text{,}
\label{34A} \\
f(\tilde{z}) &=&-5(1+\tilde{z}^{2})\alpha -\alpha \left( 49\alpha
^{2}-48\beta \right) +(1+\tilde{z})\left( 17\alpha ^{2}-7\beta \right) 
\notag \\
&&+\frac{(1+\tilde{z})\alpha ^{6}-5(1+\tilde{z})\alpha ^{4}\beta +\alpha
^{5}\beta +6(1+\tilde{z})\alpha ^{2}\beta ^{2}-4\alpha ^{3}\beta ^{2}-(1+ 
\tilde{z})\beta ^{3}+3\alpha \beta ^{3}}{\left( \alpha ^{2}-4\beta \right)
\left( (1+\tilde{z})^{2}+(1+\tilde{z})\alpha +\beta \right) }  \notag \\
&&+\frac{ 
\begin{array}{c}
128(1+\tilde{z})\alpha ^{6}-64\alpha ^{7}-720(1+\tilde{z})\alpha ^{4}\beta
+576\alpha ^{5}\beta +864(1+\tilde{z})\alpha ^{2}\beta ^{2} \\ 
-1512\alpha ^{3}\beta ^{2}-135(1+\tilde{z})\beta ^{3}+918\alpha \beta ^{3}%
\end{array}%
}{\left( \alpha ^{2}-4\beta \right) \left( (1+\tilde{z})^{2}+2\alpha (1+ 
\tilde{z})+3\beta \right) }  \notag \\
&&+\frac{ 
\begin{array}{c}
128(1+\tilde{z})\alpha ^{8}-960(1+\tilde{z})\alpha ^{6}\beta +192\alpha
^{7}\beta +2160(1+\tilde{z})\alpha ^{4}\beta ^{2}-1296\alpha ^{5}\beta ^{2}
\\ 
-1512(1+\tilde{z})\alpha ^{2}\beta ^{3}+2376\alpha ^{3}\beta ^{3}+162(1+ 
\tilde{z})\beta ^{4}-1053\alpha \beta ^{4}%
\end{array}%
}{\left( \alpha ^{2}-4\beta \right) \left( (1+\tilde{z})^{2}+2\alpha (1+ 
\tilde{z})+3\beta \right) ^{2}}\text{,}  \label{35}
\end{eqnarray}%
such that $\left[ 1+f(0)\right] A(\alpha ,\beta )\equiv 1$ where $f(\tilde{z}%
=0)\equiv f(0)$. Let us now define the effective fractional density
parameters, $\Omega _{Meff}^{\left( 0\right) }\equiv A$ and $\Omega
_{x}^{\left( 0\right) }\equiv Af(0)$ which are functions of $\alpha $, $%
\beta $ only\footnote{%
Let us note that both the (effective) dimensional density parameters $\Omega
_{Meff}$, $\Omega _{x}$ are defined in the Jordan frame though we do not put
tilde over them.}. Friedmann equation in Jordan-frame then takes the
convenient form 
\begin{equation}
{\tilde{H}^{2}}=\tilde{H}_{0}^{2}\left[ \Omega _{Meff}^{\left( 0\right) }(1+%
\tilde{z})^{3}+\Omega _{x}^{\left( 0\right) }F(\tilde{z})\right] \text{,}
\label{36}
\end{equation}%
where $F(\tilde{z})\equiv {f(\tilde{z})}/{f(0)}$. The first term in Eq. (\ref%
{36}) is the effective fractional matter density for cold matter whereas the
second term can be treated as the fractional energy density parameter of a
hypothetical fluid ($x$-fluid). For a suitable choice of numerical values of 
$\alpha $ and $\beta $ ($viz.$ $\alpha =-0.0655$ and $\beta =-0.0973$), we
have the estimates, $\Omega _{Meff}^{\left( 0\right) }=\left\{ (1+\alpha
+\beta )(1+2\alpha +3\beta )^{2}\right\} \simeq 0.2789$ and $\Omega
_{x}^{\left( 0\right) }=\left\{ (1+\alpha +\beta )(1+2\alpha +3\beta
)^{2}\right\} f(0)\simeq 0.7211$ as expected. Thus, we have succeeded to
cast the Friedmann equation in the Jordan frame in the standard form with
effective matter density, exotic fluid density. It will now be straight
forward to analyze the observational constraints on model parameters using
Eq. (\ref{36}).\newline

\section{Simple parametrization and sudden future singularity in Jordan frame%
}

\begin{figure}[tbph]
\centering
\includegraphics[width=0.50\textwidth]{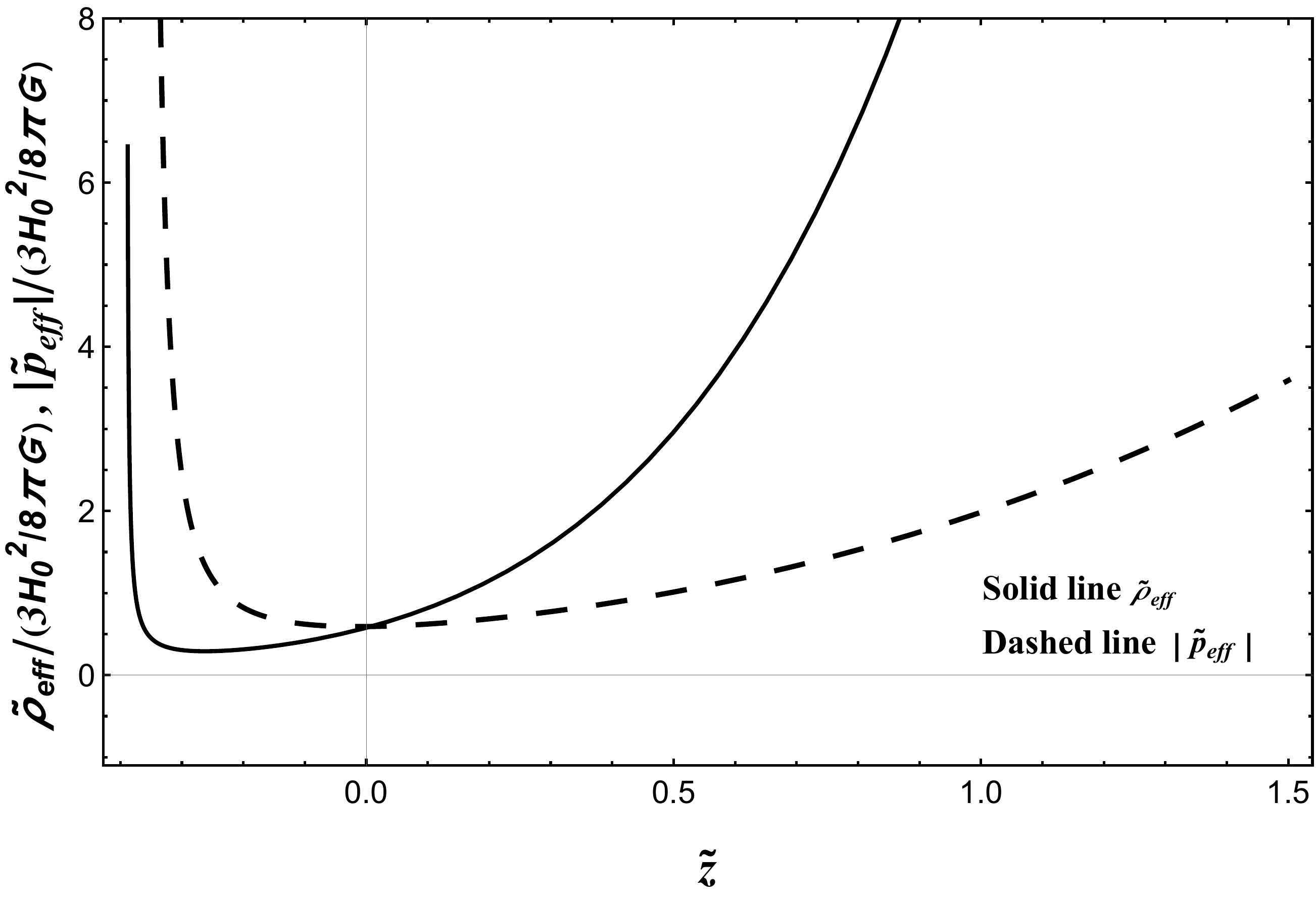}
\caption{Figure shows the behavior of effective energy density (solid line)
and effective pressure (dashed line) versus the redshift in Jordan frame.
Both the quantities diverge near $\tilde{z}=\tilde{z_{s}}=-0.3902$. This
figure also shows that effective pressure diverges faster than the effective
energy density near singularity which is consistent with the behavior of
equation of state parameter near $\tilde{z_{s}}$. Figure clearly shows the
existence of type III future singularity.}
\label{fig:rho-p}
\end{figure}

In the preceding sections, we have elaborated on the details of the scenario
based upon coupling between the known matter components in the Universe. In
particular, the model gives rise to late time cosmic acceleration in Jordan
frame. For generic values of $\alpha $ and $\beta $, Universe evolves to
deep phantom region even if the behavior is non-phantom at present. To this
effect, we quote the expressions for the effective density, pressure and
equation of state parameter (rewriting it in a slightly different form)
versus the redshift in Jordan frame in Appandix-1. Indeed, the denominator
of $\tilde{w}_{eff}$ in Eq. (\ref{w}) or Eq. (\ref{w1}) has four real roots
which are negative for generic values of $\alpha $ \& $\beta $. For the
above choice of numerical values of the parameters ($\alpha =-0.0655$ and $%
\beta =-0.0973$), the highest root is $\tilde{z_{1}}\simeq -0.3902$. The
highest root is important as the evolution would terminate there. This
expresses a future singularity which is reached for a finite value of the
redshift $\tilde{z}=\tilde{z_{s}}\equiv \tilde{z_{1}}$ ($\simeq -0.3902$) in
future. The effective equation of state diverges to minus infinity at the
quoted redshift. The effective matter density (Hubble parameter)\footnote{%
The effective matter density can be cast as effective matter density for
cold matter plus matter density of an exotic fluid, the latter diverges as $%
\tilde{z}\rightarrow \tilde{z_{s}}$} in the Jordan frame also diverge there.
Since the effective equation of state diverges, the effective pressure
should also diverge. It is clear from the structure of Eqs. (\ref{rhoeff}), (%
\ref{peff}) and (\ref{w1}) that they have the common roots. Secondly, $%
\left\vert \tilde{p}_{eff}\right\vert $ approaches infinity faster than $%
\tilde{\rho}_{eff}$ as $\tilde{z}$ approaches $\tilde{z_{s}}$, which is
consistent with the behavior of equation of state parameter given by Eq. (%
\ref{w}) or (\ref{w1}), see Fig. \ref{fig:rho-p}. Hence, the scenario under
consideration with (\ref{25}), is plagued with sudden future singularity of
type III. In what follows, we show that the sudden future singularity is not
the generic feature of the model but the artifact of simple parametrization.

\subsection{Delaying the singularity to infinite future}

In what follows, we show that by using an alternative parametrization, the
sudden future singularity can be pushed to infinite future. To this effect,
let us parametrize $a$ in terms of physical scale factor $\tilde{a}$
containing an exponential expression of the form 
\begin{equation}
a(\tilde{a})=\tilde{a}e^{\alpha \tilde{a}}\text{,}  \label{E1}
\end{equation}%
with a single model parameter $\alpha $. Expression (\ref{E1}) mimics (\ref%
{25}) for $\tilde{a}\lesssim 1$ with $\beta =\alpha ^{2}/2$ but can
crucially modify the future evolution.

\begin{figure}[tbph]
\centering
\includegraphics[width=0.55\textwidth]{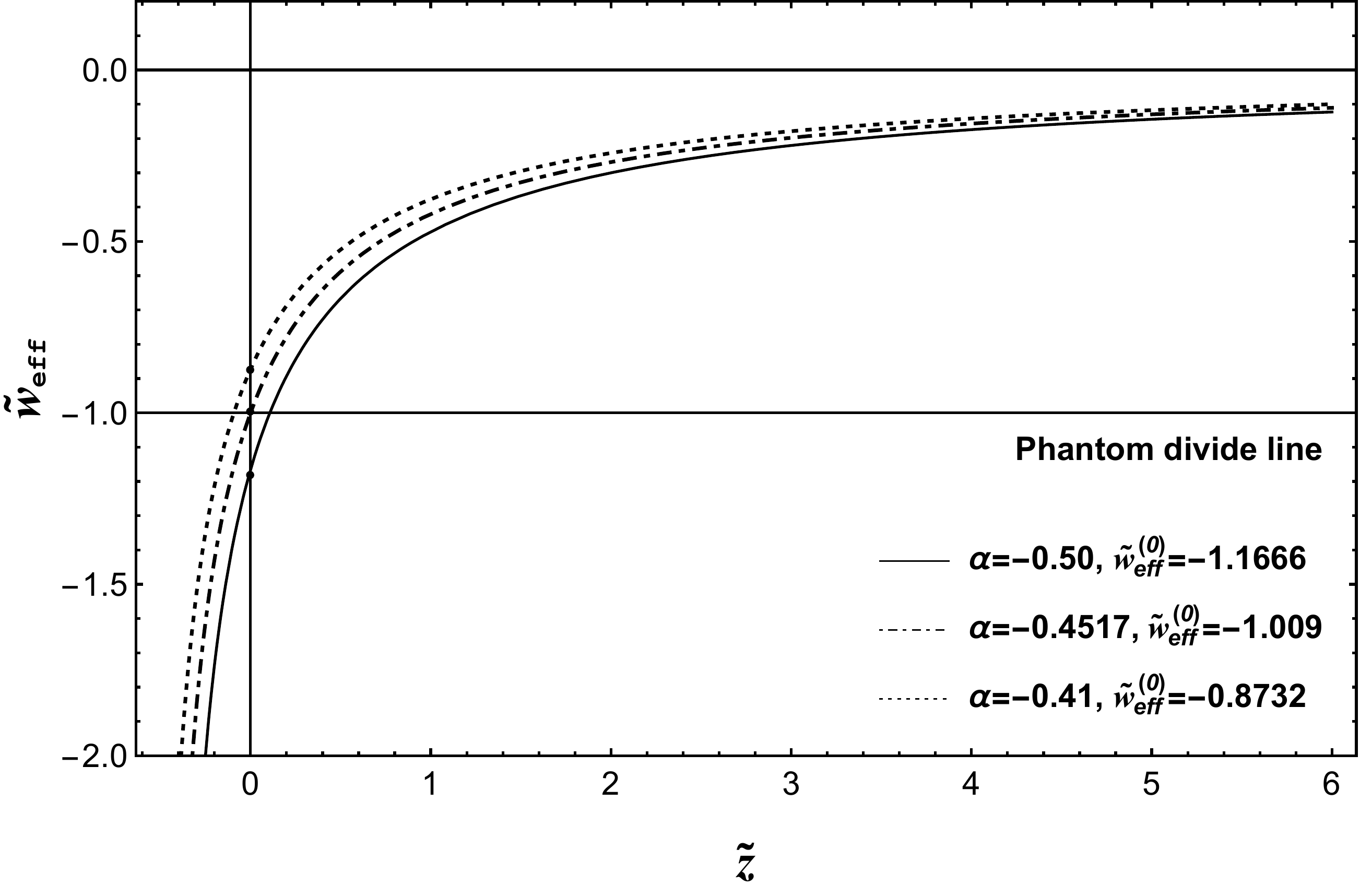}
\caption{Plot of effective equation of state parameter (\protect\ref{E4})
for different values of $\protect\alpha $ corresponding to phantom
(non-phantom) behavior at the present epoch.}
\label{fig:exp-w}
\end{figure}

In this case, the deceleration parameter in Jordan-frame is given by%
\begin{equation}
\tilde{q}=-\frac{\tilde{a}\ddot{\tilde{a}}}{\dot{\tilde{a}}^{2}}=\frac{
\left( 1+\alpha \tilde{a}\right) ^{2}+2\tilde{a}(2\alpha +\alpha ^{2}\tilde{%
a })}{2\left( 1+\alpha \tilde{a}\right) }\text{.}  \label{E3}
\end{equation}
The effective equation of state in Jordan frame is then obtained as

\begin{equation}
\tilde{w}_{eff}(\tilde{z})=\frac{5\alpha (1+\tilde{z})+3\alpha ^{2}}{3(1+ 
\tilde{z})\left[ (1+\tilde{z})+\alpha \right] }\text{ .}  \label{E4}
\end{equation}%
Unlike, (\ref{25}), we have only one model parameter in this expression of
effective equation of state parameter induced by coupling between the known
components of matter. In this case, 
\begin{equation}
\tilde{w}_{eff}^{(0)}=\frac{5\alpha +3\alpha ^{2}}{3(1+\alpha )}\text{,}
\label{E5}
\end{equation}%
which can have a desired value for a chosen negative values of $\alpha $;
for instance, $\tilde{w}_{eff}^{(0)}\simeq -0.98$ for $\alpha \simeq -0.44$.
We have shown a plot (see Fig. \ref{fig:exp-w}) of the effective equation of
state parameter (\ref{E4}) for different values of $\alpha $ corresponding
to phantom (non-phantom) behavior at the present epoch. Let us note that
unlike the previous case (\ref{w}), the denominator in (\ref{E4}) does not
vanish for any negative values of the redshift ($\tilde{z}>-1$). The
underlying system would always evolve to phantom in future without hitting
the singularity. In this case, the singularity would occur at $\tilde{a_{s}}%
=\infty $. Clearly, sudden future singularity is an artifact of
extrapolation of (\ref{25}) beyond the present epoch where, strictly
speaking, the latter is not valid. It is really interesting that one
parameter family (\ref{E1}) gives rise to late time acceleration without
sudden future singularity.

\section{Observational constrains}

In the preceding sections, we have analyzed different cosmological aspects
of the scenario based upon coupling between normal components of matter. We
now proceed to constrain the model parameters $\alpha $ \& $\beta $ using
H(z), SNIa and BAO data. The model parameters $\alpha $ \& $\beta $ are
constrained by employing the $\chi ^{2}$ analysis. The maximum likelihood
method is used and the total likelihood for $\alpha $ \& $\beta $ as the
product of individual likelihood for different datasets is obtained. For the
joint data, the total likelihood function is written as 
\begin{equation}
\mathcal{L}_{tot}(\alpha ,\beta )=e^{-\frac{\chi _{tot}^{2}(\alpha ,\beta )}{
2}}\text{,}  \label{eq:likli}
\end{equation}%
where we have 
\begin{equation}
\chi _{\mathrm{tot}}^{2}=\chi _{\mathrm{Hub}}^{2}+\chi _{\mathrm{SN}%
}^{2}+\chi _{\mathrm{BAO}}^{2}\,,  \label{eq:chi}
\end{equation}%
and is related to the Hubble ($H(z)$) dataset, the Supernovae of Type Ia
(SNIa) and the Baryon Acoustic Oscillation (BAO) data. The best fit value of
the model parameters $\alpha $ \& $\beta $ is obtained by minimizing the
total chi-square $\chi _{\mathrm{tot}}^{2}$ with respect to $\alpha $ \& $%
\beta $. As usual, the likelihood contours at 1$\sigma $ and 2$\sigma $
confidence level are $2.3$ and $6.17$, respectively, in the two dimensional
plane. The basic tools and the reference data used here for the data
analysis are given in appendix-2.

\begin{figure}[tbph]
\begin{center}
\begin{tabular}{ccc}
{\includegraphics[width=2.5in,height=2.4in,angle=0]{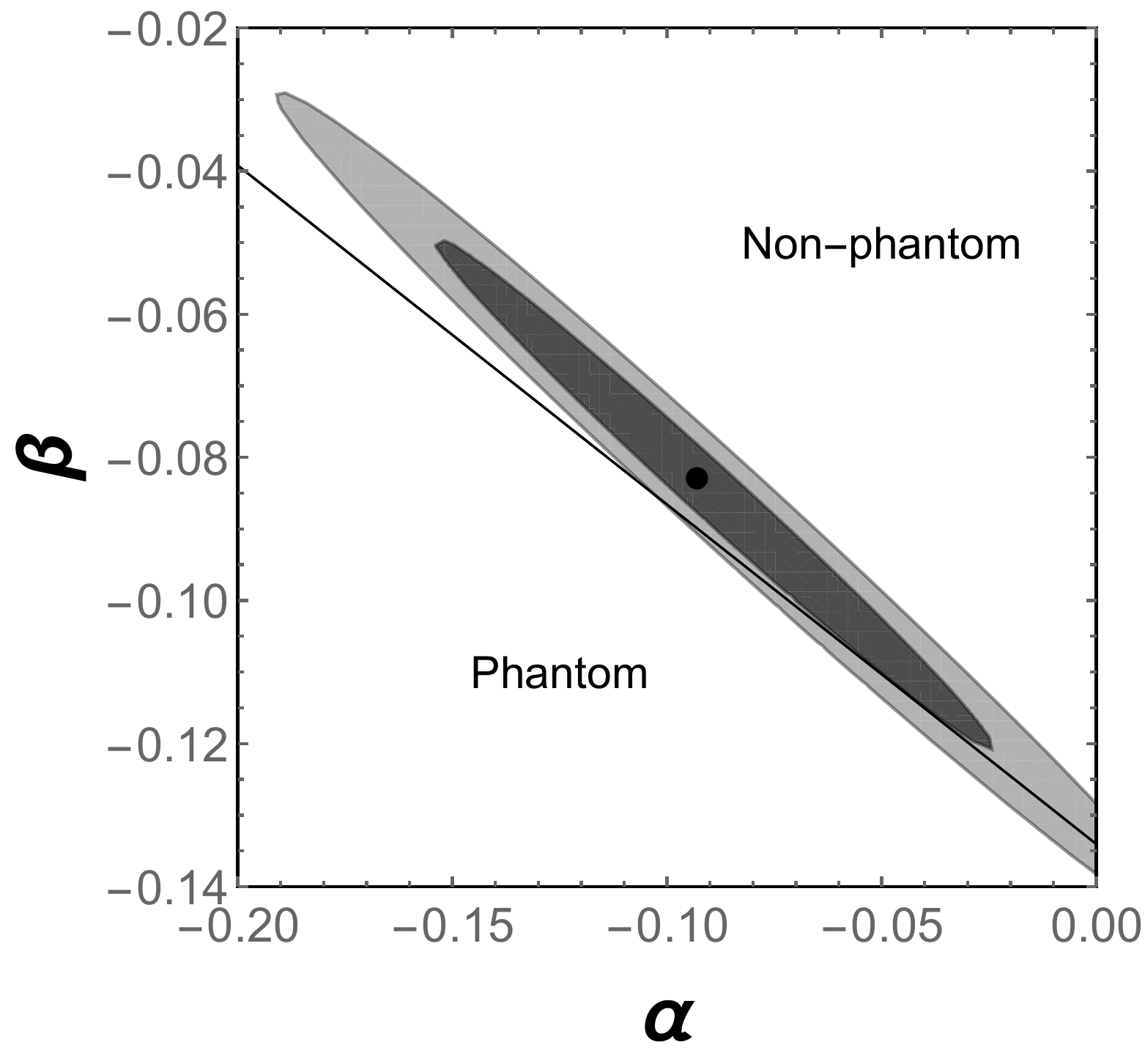}} & {\ %
\includegraphics[width=2.5in,height=2.4in,angle=0]{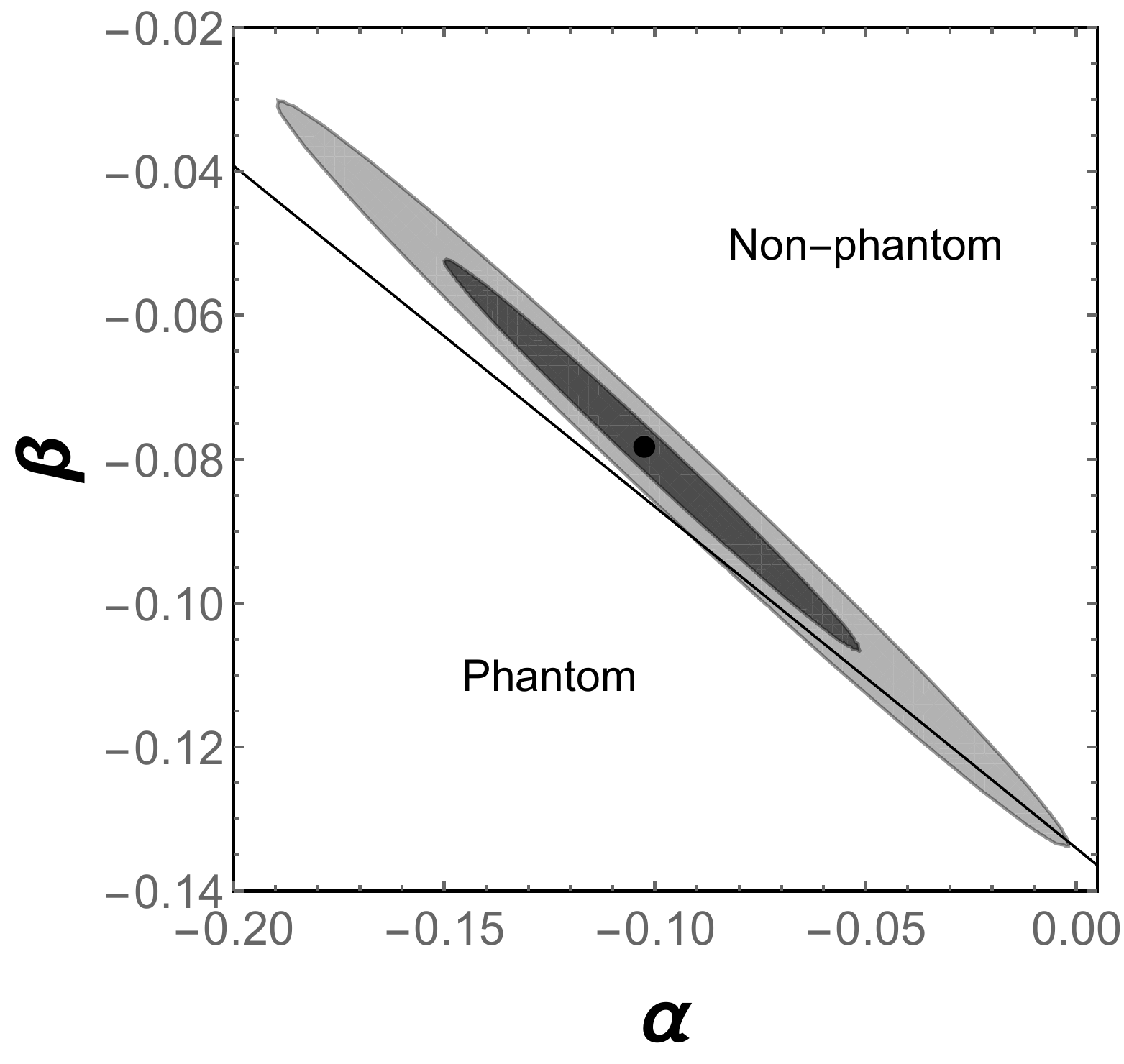}} & 
\end{tabular}%
\end{center}
\caption{The figure shows the 1$\protect\sigma $ (dark shaded) and 2$\protect%
\sigma $ (light shaded) likelihood contours in $\protect\alpha \protect\beta %
-$plane. The left panel corresponds to SN+BAO whereas right panel is for
Hubble+SN+BAO. The black dots represent the best fit value of the model
parameters which are found to be $\protect\alpha =-0.092954$, $\protect\beta %
=-0.083058$ (left panel) and $\protect\alpha =-0.102681$, $\protect\beta %
=-0.078347$ (right panel). The dashed line shown in both the plots is the $%
\tilde{w}_{eff}^{(0)}=-1$ line which intersects the contours in 2$\protect%
\sigma $ region and separates the phantom and non phantom region. We get a
narrow strip from both the SN+BAO and Hubble+SN+BAO constraints for the
allowed values of $\protect\alpha $ \& $\protect\beta $ for phantom
evolution. }
\label{fig:cont}
\end{figure}

In figure \ref{fig:cont}, we have shown the results based upon $SNIa+BAO$
data and the combined data set, $H(z)+SNIa+BAO$. For the both the data sets,
there is a region of parameter space where the model is close to $\Lambda $%
CDM. The region with relatively larger values of $\alpha $ and smaller
values of $\beta $ (numerically) is some what away from $\Lambda $CDM model.
The best fit values of cosmological parameters are close to each other for
the two data sets. For instance, $\Omega _{Meff}^{(0)}\simeq 0.26$, thereby,
the model favors relatively lower values of matter density parameter
compared to $\Lambda $CDM. The present values of other cosmological
parameters are listed in the following table \ref{Cons-tab}.

\begin{table}[tbph]
\begin{center}
\begin{tabular}{|c|c|c|}
\hline
$Data$\ $sets$ & $SNIa+BAO$ & $H(z)+SNIa+BAO$ \\ \hline
$Best$ $Fit$ $Values$ & 
\begin{tabular}{l}
$\alpha =-0.09295$ \\ 
$\beta =-0.08305$%
\end{tabular}
& 
\begin{tabular}{l}
$\alpha =-0.10268$ \\ 
$\beta =-0.07834$%
\end{tabular}
\\ \hline
$\tilde{w}_{eff}^{(0)}$ & $-0.91218$ & $-0.91015$ \\ \hline
$\Omega _{Meff}^{(0)}$ & $0.26299$ & $0.25648$ \\ \hline
$\Omega _{X}^{(0)}$ & $0.73701$ & $0.74352$ \\ \hline
$\tilde{t}_{0}$\ $(Gyr)$ & $14.054$ & $14.102$ \\ \hline
\end{tabular}%
\end{center}
\caption{Table indicates the best fit values of $\protect\alpha $, $\protect%
\beta $ obtained using $SNIa+BAO$ and $H(z)+SNIa+BAO$ data sets with
corresponding values of other cosmological parameters.}
\label{Cons-tab}
\end{table}

\section{Conclusions and outlook}

In this paper we further investigated the proposal of Ref. \cite{KHOURY2016}
which is an alternative to dark energy and large scale modification of
gravity associated with \textit{extra degrees of freedom coupled to matter}.
In this picture, the late time cosmic acceleration is sourced by a general
type of coupling between dark matter and baryonic matter. In Einstein frame,
both the components individually do not follow standard conservation due to
coupling between them though the total energy density does and should.
Assuming both the components to be pressure-less, their total energy density
redshifts as usual in the Einstein frame. Clearly, there is no acceleration
in this frame in the framework under consideration.

We should emphasize that coupling is defined in the Einstein frame only. As
for the, Jordan frame, both the matter components adhere to standard
conservation, however, the Einstein Hilbert action and matter action are
modified, see Eq. (\ref{riccij}) in Appendix-1. This modification of gravity
is not endowed with extra degrees of freedom as there is non in the Einstein
frame where the Lagrangian is diagonalized. The transformation of metric
contains two functions $R$ and $Q$ to be chosen keeping in mind
phenomenology, namely, in high density regime they should reduce to unity
leaving the thermal history and local physics intact. Secondly, at late
stages, the choice of metric functions should give rise to accelerating
Universe. The choice, $Q=1$ corresponds to maximally dis-conformal
transformation. In this case, $R$ being concave up and fast growing at late
times can give rise to late time cosmic acceleration in the Jordan frame.

A comment about the conformal coupling which corresponds to $Q=R$ is in
order. The conformal choice though disfavored from stability criteria but
not prohibited \cite{KHOURY2016}. It simply requires the fine tuning of the
metric functions, thereby, in principle, late time cosmic acceleration is
possible in both the conformal and disformal cases.

Using the two parameter functional form for $a(\tilde{a})$, we derived
expression for the ratio, $\tilde{H}(\tilde{z})/\tilde{H}(0)$, which is a
rational expression in $\alpha ,$ $\beta $ $\&$ $\tilde{z}$. We show that
the age computed using the rational expression in the Jordan frame is
consistent with observation provided we choose the couplings in accordance
with the observed value of deceleration parameter. In the Einstein frame, we
have standard FRW Universe inhabited with cold matter giving rise to $%
t_{0}=2/3H_{0}$ as expected. In order to reconcile the scenario with the
standard lore, we transformed the ratio, $\tilde{H}(\tilde{z})/\tilde{H}(0)$%
, to a suggestive form isolating the cold matter like term proportional to $%
(1+\tilde{z})^{3}$ and attributing the rest to a hypothetical
\textquotedblleft $x$\textquotedblright -fluid. The so defined effective
fractional energy densities $\Omega _{Meff}^{(0)}$ $\&$ $\Omega _{x}^{(0)}$
are functions of $\alpha $ and $\beta $ alone. The best fit values for the
couplings allows us to reconcile with the standard lore with cold matter and
an exotic fluid with large negative pressure giving rise to $\Omega
_{Meff}^{(0)}\simeq 0.26299$ $\&$ $\Omega _{x}^{(0)}\simeq 0.73701$ (for $%
SNIa+BAO$ data) and $\Omega _{Meff}^{(0)}\simeq 0.25648$ $\&$ $\Omega
_{x}^{(0)}\simeq 0.74352$ (for $H(z)+SNIa+BAO$ data).

The underlying picture should be contrasted with large scale modification of
gravity due to extra degree(s) of freedom where both the frames are related
to each other by conformal transformation. In that case, proper screening of
the extra degrees does not leave any scope for late time acceleration$-$ 
\textit{acceleration in Jordan frame and no acceleration in Einstein frame}.
Indeed, if one screens out the local effect of the extra degrees of freedom,
conformal transformation from Jordan to Einstein frame fails to remove
acceleration completely, thereby, acceleration is not due to modification of
gravity. This result applies to any scheme of large scale modification
caused by massive extra degrees of freedom coupled to matter such as $f(R)$ theories \cite{EXTENDED, CAPOZZI}. It is
remarkable that in the framework, under consideration, acceleration is
completely removed by disformal (conformal) transformation, though one
adheres to disformal transformations only in order to avoid fine tuning
related to instability. Hence, the scenario based upon disformal coupling
between dark matter and baryonic matter, represents the true large scale
modification of gravity as the underlying cause of late time cosmic
acceleration.

Let us also mention that the scenario admits phantom as well as non-phantom
behavior provided that we make suitable choice of parameters $\alpha $ $\&$ $%
\beta $. Indeed, as shown, in Fig. \ref{fig:cont}, there is narrow region
below the phantom divide line in the $2\sigma $ contour allowed by the
combined data . In this framework, if simple parametrization used, Universe
inevitable evolves to phantom even if the present behavior is set to be
non-phantom. In case of the parametrization (\ref{25}), the Hubble
parameter, effective matter density and effective pressure diverge at a
finite value of the red-shift, $\tilde{z_{s}}\simeq -0.39932$ (for $SNIa+BAO$
data) and $\tilde{z_{s}}\simeq -0.40$ (for $H(z)+SNIa+BAO$ data) \textit{a la%
} a type III sudden future singularity. This is, however, not the generic
feature of the model but rather the non-judicial use of (\ref{25}) beyond
present epoch in future. Indeed, we have shown that replacing (\ref{25}) by
an alternative parametrization (\ref{E1}), the sudden future singularity can
be delayed to infinite future leaving intact the dynamics from past to the
present epoch. 

In our opinion, the scenario is of great interest and deserves further
investigations both at the background as well as the level of perturbations.

\section{Acknowledgement}

We thank Subhadip Mitra, V. Sahni and Ujjaini Alam for useful discussions.
We are indebted to M. Shahalam and Safia Ahmad for help in data analysis. S.
K. J. Pacif wishes to thank the National Board of Higher Mathematics (NBHM),
Department of Atomic Energy (DAE), Govt. of India for financial support
through the post-doctoral research fellowship. The work of A. Wang is partly
supported by the National Natural Science Foundation of China (NNSFC) with
the Grant Nos.: 11375153 and 11173021.

\section{Appendix-1: Ricci scalar in Jordan frame}

In the present scenario, dark matter and baryonic matter are disformally
coupled in Einstein frame such that the individual components do not
conserve on their own but standard conservation applies to both of them take
together. As a result, Universe is matter dominated after radiation matter
equality, thereby decelerating, in Einstein frame. Let us emphasize again
that coupling is defined in Einstein frame only. We can transform to Jordan
frame where the matter components adhere to standard conservation separately
but dynamics becomes complicated. In particular, the Einstein-Hilbert action
transforms, In the flat FRW background, the Jordan frame metric is $\tilde{g}%
_{\mu \nu }$ with the coupling functions $Q=Q(a)$ and $R=R(a)$ is

\begin{equation*}
\tilde{g}_{\mu \nu }=diag(-Q^{2},R^{2}a^{2},R^{2}a^{2},R^{2}a^{2}).
\end{equation*}%
In this background, the Ricci scalar in Jordan frame ($\mathcal{\tilde{R}}%
_{J}$), is given by


\begin{equation}
\mathcal{\tilde{R}}_{J}=Q^{-2}\mathcal{R}_{E}+6Q^{-2}\left[ \ddot{a}\frac{
R^{^{\prime }}}{R}+4\frac{\dot{a}^{2}}{a}\frac{R^{^{\prime }}}{R}+\dot{a}^{2}%
\frac{R^{^{\prime }2}}{R^{2}}+\dot{a}^{2}\frac{R^{^{\prime \prime }}}{R}-%
\dot{a}^{2}\frac{Q^{^{\prime }}}{Q}\frac{R^{^{\prime }}}{R}-\frac{\dot{a}^{2}%
}{a}\frac{Q^{^{\prime }}}{Q}\right] ,  \label{riccij}
\end{equation}%
where dot and dash designate derivatives with respect to time cosmic time
\textquotedblleft $t$\textquotedblright\ and scale factor \textquotedblleft $%
a$\textquotedblright\ respectively; the Einstein frame Ricci scalar is given
by, $\mathcal{R}_{E}=\frac{6}{a^{2}}\left( a\ddot{a}+\dot{a}^{2}\right) $.
For maximally disformal case, $Q=1$, $R=R(a)$, we have,%
\begin{equation}
\mathcal{\tilde{R}}_{J}=\mathcal{R}_{E}+6\left[ \ddot{a}\frac{R^{^{\prime }}%
}{R}+4\frac{\dot{a}^{2}}{a}\frac{R^{^{\prime }}}{R}+\dot{a}^{2}\frac{
R^{^{\prime }2}}{R^{2}}+\dot{a}^{2}\frac{R^{^{\prime \prime }}}{R}\right] .
\label{ricci}
\end{equation}%
which reduces to Ricci scalar in Einstein frame at early times as $R$ turns
constant there. For the parametrization $a(\tilde{a})=\tilde{a}+\alpha 
\tilde{a}^{2}+\beta \tilde{a}^{3}$, $\mathcal{\tilde{R}}_{J}$ can be recast
in terms of redshift ($\tilde{z}=\frac{1}{\tilde{a}}-1$) as

\begin{equation*}
\mathcal{\tilde{R}}_{J}=\frac{\left[ 
\begin{array}{c}
\left\{ 3H_{0}^{2}(1+\alpha +\beta )(1+2\alpha +3\beta )^{2}(1+\tilde{z}
)^{9}\right\} \\ 
\times \left\{ 
\begin{array}{c}
(1+\tilde{z})^{3}\left[ (1+\tilde{z})^{3}-\alpha (1+\tilde{z})^{2}+2\alpha
^{2}(1+\tilde{z})(-3+4\tilde{z})+4\alpha ^{3}(-1+3\tilde{z})\right] \\ 
+\beta (1+\tilde{z})^{2}\left[ (1+\tilde{z})^{2}(-1+4\tilde{z})+6\alpha (1+ 
\tilde{z})(-1+8\tilde{z})+2\alpha ^{2}(-3+38\tilde{z})\right] \\ 
+\beta ^{2}(1+\tilde{z})\left[ 7(1+\alpha )+\tilde{z}(63+144\alpha +56\tilde{
z})\right] \\ 
+3\beta ^{3}(3+28\tilde{z})%
\end{array}
\right\}%
\end{array}
\right] }{\left[ \left\{ (1+\tilde{z})(1+\tilde{z}+\alpha )+\beta \right\}
^{3}\left\{ (1+\tilde{z})(1+\tilde{z}+2\alpha )+3\beta \right\} ^{3}\right] }%
\text{.}
\end{equation*}%
The denominator of $\mathcal{\tilde{R}}_{J}$ has the same structure present
in the equation of state parameter, effective pressure and effective energy
density. Clearly $\mathcal{\tilde{R}}_{J}$ diverges as $\tilde{z}\rightarrow 
\tilde{z_{s}}$. Indeed, there are four roots,

\begin{eqnarray*}
\tilde{z} &=&\frac{1}{2}\left\{ -2-\alpha -\sqrt{\alpha ^{2}-4\beta }
\right\} ,\frac{1}{2}\left\{ -2-\alpha +\sqrt{\alpha ^{2}-4\beta }\right\} ,
\\
&&\left\{ -1-\alpha -\sqrt{\alpha ^{2}-3\beta }\right\} ,\left\{ -1-\alpha + 
\sqrt{\alpha ^{2}-3\beta }\right\} \text{.}
\end{eqnarray*}

For $\alpha =-0.0655,$ $\beta =-0.0973$, we have $\tilde{z}\simeq
-0.3902,-0.6536,-1.2806,-1.4781$. At $\tilde{z}=\tilde{z_{s}}\simeq -0.3902$
around which $\mathcal{\tilde{R}}_{J}$ diverges.

Let us also quote the expressions for the effective pressure, matter density
and equation of state parameter in Jordan frame required for the
classification of singularity 

\begin{eqnarray}
\tilde{\rho}_{eff}(\tilde{z}) &=&\frac{3\tilde{H}_{0}^{2}}{8\pi G}\frac{
(1+\alpha +\beta )(1+2\alpha +3\beta )^{2}\left( 1+\tilde{z}\right) ^{9}}{ %
\left[ \left( 1+\tilde{z}\right) ^{2}+\alpha \left( 1+\tilde{z}\right)
+\beta \right] \left[ \left( 1+\tilde{z}\right) ^{2}+2\alpha \left( 1+\tilde{
z}\right) +3\beta \right] ^{2}}\text{,}  \label{rhoeff} \\
\tilde{p}_{eff}(\tilde{z}) &=&\frac{\tilde{H}_{0}^{2}}{8\pi G}\left[ \frac{
(1+\alpha +\beta )(1+2\alpha +3\beta )^{2}\left( 1+\tilde{z}\right) ^{9}}{ %
\left[ \left( 1+\tilde{z}\right) ^{2}+\alpha \left( 1+\tilde{z}\right)
+\beta \right] \left[ \left( 1+\tilde{z}\right) ^{2}+2\alpha \left( 1+\tilde{
z}\right) +3\beta \right] ^{2}}\right]  \notag \\
&&\times \left[ -1+\left( \frac{\left( 1+\tilde{z}\right) ^{2}+2\alpha
\left( 1+\tilde{z}\right) +3\beta }{\left( 1+\tilde{z}\right) ^{2}+\alpha
\left( 1+\tilde{z}\right) +\beta }\right) +\left( \frac{4(\left( 1+\tilde{z}
\right) +3\beta )}{\left( 1+\tilde{z}\right) ^{2}+2\alpha \left( 1+\tilde{z}
\right) +3\beta }\right) \right] \text{,}  \label{peff} \\
\tilde{w}_{eff}(\tilde{z}) &=&\frac{1}{3}\left[ -1+\left( \frac{\left( 1+ 
\tilde{z}\right) ^{2}+2\alpha \left( 1+\tilde{z}\right) +3\beta }{\left( 1+ 
\tilde{z}\right) ^{2}+\alpha \left( 1+\tilde{z}\right) +\beta }\right)
+\left( \frac{4(\left( 1+\tilde{z}\right) +3\beta )}{\left( 1+\tilde{z}
\right) ^{2}+2\alpha \left( 1+\tilde{z}\right) +3\beta }\right) \right] 
\text{.}  \label{w1}
\end{eqnarray}


\section{Appendix-2: Tools for Data Analysis}

For analysis with H(z) data, We have used the compiled dataset used by
Farooq and Ratra \cite{Farooq:2013hq} of $28$ $H(z)$ data points in the
redshift range $0.07 \leq z \leq 2.3$. We use $H_{0}=67.3\pm 1.2~Km/S/Mpc$
to complete the dataset \cite{planck}. We apply the data to the model with
normalized Hubble parameter, $h=H/H_{0}$. In this case, the $\chi ^{2}$ is
defined as 
\begin{equation}
\chi _{\mathrm{Hub}}^{2}(\theta )=\sum_{i=1}^{29}\frac{\left[ h_{\mathrm{th}
}(z_{i},\theta )-h_{\mathrm{obs}}(z_{i})\right] ^{2}}{\sigma _{h}(z_{i})^{2}}%
\,,
\end{equation}
where $h_{\mathrm{obs}}$ and $h_{\mathrm{th}}$ are the observed and
theoretical values of the normalized Hubble parameter, respectively, and $%
\sigma _{h}=\left( \frac{\sigma _{H}}{H}+\frac{\sigma _{H_{0}}}{H_{0}}%
\right) h$, where $\sigma _{H}$ and $\sigma _{H_{0}}$ are the errors
associated with $H$ and ${H_{0}}$, respectively.

As we know that Type Ia supernova is considered as an ideal astronomical
object and observed as very good standard candles. It is one of the direct
probe for the cosmological expansion. For our analysis, we take $580$ data
points from Union2.1 compilation data \cite{Suzuki:2011hu}. The appropriate
quantity luminosity distance $D_{L}(z)$ can be defined as 
\begin{equation}
D_{L}(z)=(1+z)\int_{0}^{z}\frac{H_{0}dz^{\prime }}{H(z^{\prime })}.
\label{D2}
\end{equation}
The distance modulus $\mu (z)$ is the observed quantity and is related to $%
D_{L}(z)$ as $\mu (z)=m-M=5\log D_{L}(z)+\mu _{0}$, where $m$ and $M$ are
the apparent and absolute magnitudes of the Supernovae and $\mu _{0}=5\log
\left( \frac{H_{0}^{-1}}{\mathrm{Mpc}}\right) +25$ is the nuisance parameter
that should be marginalized. The corresponding $\chi ^{2}$ is given by 
\begin{equation}
\chi _{\mathrm{SN}}^{2}(\mu _{0},\theta )=\sum_{i=1}^{580}\frac{\left[ \mu
_{th}(z_{i},\mu _{0},\theta )-\mu _{obs}(z_{i})\right] ^{2}}{\sigma _{\mu
}(z_{i})^{2}}\,,
\end{equation}
where $\mu _{th}$, $\mu _{obs}$ and $\sigma _{\mu }$ represent the
theoretical, observed distance modulus and uncertainty in the distance
modulus, respectively. The $\theta $ is an arbitrary parameter of the
corresponding model. After marginalization of $\mu _{0}$ \cite{Lazkoz05},
one obtains 
\begin{equation}
\chi _{\mathrm{SN}}^{2}(\theta )=A(\theta )-\frac{B(\theta )^{2}}{C(\theta )}%
\,,
\end{equation}
where, 
\begin{align}
& A(\theta )=\sum_{i=1}^{580}\frac{\left[ \mu _{th}(z_{i},\mu _{0}=0,\theta
)-\mu _{obs}(z_{i})\right] ^{2}}{\sigma _{\mu }(z_{i})^{2}}\,, \\
& B(\theta )=\sum_{i=1}^{580}\frac{\mu _{th}(z_{i},\mu _{0}=0,\theta )-\mu
_{obs}(z_{i})}{\sigma _{\mu }(z_{i})^{2}}\,, \\
& C(\theta )=\sum_{i=1}^{580}\frac{1}{\sigma _{\mu }(z_{i})^{2}}\,.
\end{align}

Finally, we worked out with $BAO$ data of $d_{A}(z_{\star })/D_{V}(Z_{BAO})$ 
\cite{Blake:2011en, Percival:2009xn, Beutler:2011hx, Jarosik:2010iu,
Eisenstein:2005su, Giostri:2012ek}, where $z_{\star }\approx 1091$ is the
decoupling time, $d_{A}(z)$ is the co-moving angular-diameter distance and $%
D_{V}(z)=\left( d_{A}(z)^{2}z/H(z)\right) ^{1/3}$ is the dilation scale. The
required data is presented in Table \ref{BAOt}. The corresponding $\chi _{ 
\mathrm{BAO}}^{2}$ is given by \cite{Giostri:2012ek}: 
\begin{equation}
\chi _{\mathrm{BAO}}^{2}=Y^{T}C^{-1}Y\,,
\end{equation}%
where, 
\begin{equation}
Y=\left( 
\begin{array}{c}
\frac{d_{A}(z_{\star })}{D_{V}(0.106)}-30.95 \\ 
\frac{d_{A}(z_{\star })}{D_{V}(0.2)}-17.55 \\ 
\frac{d_{A}(z_{\star })}{D_{V}(0.35)}-10.11 \\ 
\frac{d_{A}(z_{\star })}{D_{V}(0.44)}-8.44 \\ 
\frac{d_{A}(z_{\star })}{D_{V}(0.6)}-6.69 \\ 
\frac{d_{A}(z_{\star })}{D_{V}(0.73)}-5.45%
\end{array}%
\right) \,,
\end{equation}%
and $C^{-1}$ is the inverse covariance matrix given by \cite{Giostri:2012ek}.

\begin{center}
\begin{tabular}{|c|c|}
\hline
$z_{BAO}$ & $\frac{d_{A}(z_{\star })}{D_{V}(Z_{BAO})}$ \\ \hline
0.106 & $30.95\pm 1.46$ \\ \hline
0.2 & $17.55\pm 0.60$ \\ \hline
0.35 & $10.11\pm 0.37$ \\ \hline
0.44 & $8.44\pm 0.67$ \\ \hline
0.6 & $6.69\pm 0.33$ \\ \hline
0.73 & $5.45\pm 0.31$ \\ \hline
\end{tabular}

Values of $\frac{d_{A}(z_{\star })}{D_{V}(Z_{BAO})}$ for distinct values of $%
z_{BAO}$. 
\begin{equation*}
C^{-1}=\left( 
\begin{array}{cccccc}
0.48435 & -0.101383 & -0.164945 & -0.0305703 & -0.097874 & -0.106738 \\ 
-0.101383 & 3.2882 & -2.45497 & -0.0787898 & -0.252254 & -0.2751 \\ 
-0.164945 & -2.45499 & 9.55916 & -0.128187 & -0.410404 & -0.447574 \\ 
-0.0305703 & -0.0787898 & -0.128187 & 2.78728 & -2.75632 & 1.16437 \\ 
-0.097874 & -0.252254 & -0.410404 & -2.75632 & 14.9245 & -7.32441 \\ 
-0.106738 & -0.2751 & -0.447574 & 1.16437 & -7.32441 & 14.5022%
\end{array}
\right) \,.
\end{equation*}
\end{center}


\begin{thebibliography}{99}
\bibitem{HZTEAM} A.~G.~Riess \textit{et al.} [Supernova Search Team], 
Astron.\ J.\ \textbf{116}, 1009 (1998) 
[astro-ph/9805201]. 

\bibitem{SCP} S.~Perlmutter \textit{et al.} [Supernova Cosmology Project
Collaboration], 
Astrophys.\ J.\ \textbf{517}, 565 (1999) 
[astro-ph/9812133]. 

\bibitem{spergel} D.~N.~Spergel \textit{et al.} [WMAP Collaboration], 
Astrophys.\ J.\ Suppl.\ \textbf{170}, 377 (2007) 
[astro-ph/0603449]. 

\bibitem{seljak} U.~Seljak \textit{et al.} [SDSS Collaboration], 
Phys.\ Rev.\ D \textbf{71}, 103515 (2005) 
[astro-ph/0407372]. 

\bibitem{DE1} E.~J.~Copeland, M.~Sami and S.~Tsujikawa, 
Int.\ J.\ Mod.\ Phys.\ D \textbf{15}, 1753 (2006) 
[hep-th/0603057]. 

\bibitem{DE2} M.~Sami, 
New Adv.\ Phys.\ \textbf{10}, 77 (2016) [arXiv:1401.7310 [physics.pop-ph]]. 

\bibitem{DE3} V.~Sahni and A.~A.~Starobinsky, 
Int.\ J.\ Mod.\ Phys.\ D \textbf{9}, 373 (2000) 
[astro-ph/9904398]. 

\bibitem{DE4} J.~Frieman, M.~Turner and D.~Huterer, 
Ann.\ Rev.\ Astron.\ Astrophys.\ \textbf{46}, 385 (2008) 
[arXiv:0803.0982 [astro-ph]]. 

\bibitem{DE5} R.~R.~Caldwell and M.~Kamionkowski, 
Ann.\ Rev.\ Nucl.\ Part.\ Sci.\ \textbf{59}, 397 (2009) 
[arXiv:0903.0866 [astro-ph.CO]]. 

\bibitem{DE6} A.~Silvestri and M.~Trodden, 
Rept.\ Prog.\ Phys.\ \textbf{72}, 096901 (2009) 
[arXiv:0904.0024 [astro-ph.CO]]. 

\bibitem{DE7} M.~Sami, 
Curr.\ Sci.\ \textbf{97}, 887 (2009) [arXiv:0904.3445 [hep-th]]. 

\bibitem{DE8} L.~Perivolaropoulos, 
AIP Conf.\ Proc.\ \textbf{848}, 698 (2006) 
[astro-ph/0601014]. 

\bibitem{DE9} J.~A.~Frieman, 
AIP Conf.\ Proc.\ \textbf{1057}, 87 (2008) 
[arXiv:0904.1832 [astro-ph.CO]]. 

\bibitem{DE10} M.~Sami, 
Lect.\ Notes Phys.\ \textbf{720}, 219 (2007). 

\bibitem{carollCC} S.~M.~Carroll, 
Living Rev.\ Rel.\ \textbf{4}, 1 (2001) 
[astro-ph/0004075]. 

\bibitem{padmaCC} T.~Padmanabhan, 
Phys.\ Rept.\ \textbf{380}, 235 (2003) 
[hep-th/0212290]. 

\bibitem{peeblesCC} P.~J.~E.~Peebles and B.~Ratra, 
Rev.\ Mod.\ Phys.\ \textbf{75}, 559 (2003) 
[astro-ph/0207347]. 

\bibitem{quint1} C.~Wetterich, 
Nucl.\ Phys.\ B \textbf{302}, 668 (1988). 

\bibitem{quint2} B.~Ratra and P.~J.~E.~Peebles, 
Phys.\ Rev.\ D \textbf{37}, 3406 (1988). 

\bibitem{quint3} R.~R.~Caldwell, R.~Dave and P.~J.~Steinhardt, 
Phys.\ Rev.\ Lett.\ \textbf{80}, 1582 (1998) 
[astro-ph/9708069]. 

\bibitem{quint4} V.~Sahni, M.~Sami and T.~Souradeep, 
Phys.\ Rev.\ D \textbf{65}, 023518 (2002) 
[gr-qc/0105121]. 

\bibitem{quint5} M.~Sami and T.~Padmanabhan, 
Phys.\ Rev.\ D \textbf{67}, 083509 (2003) 
[hep-th/0212317]. 

\bibitem{LPC1} A.~Upadhye, W.~Hu and J.~Khoury, 
Phys.\ Rev.\ Lett.\ \textbf{109}, 041301 (2012) 
[arXiv:1204.3906 [hep-ph]]. 

\bibitem{LPC2} R.~Gannouji, M.~Sami and I.~Thongkool, 
Phys.\ Lett.\ B \textbf{716}, 255 (2012) 
[arXiv:1206.3395 [hep-th]]. 

\bibitem{chamel0} D.~F.~Mota and J.~D.~Barrow, 
Phys.\ Lett.\ B \textbf{581}, 141 (2004) 
[astro-ph/0306047]. 

\bibitem{chamel1} J.~Khoury and A.~Weltman, 
Phys.\ Rev.\ Lett.\ \textbf{93}, 171104 (2004) 
[astro-ph/0309300]. 

\bibitem{chamel2} P.~Brax, C.~van de Bruck, A.~C.~Davis, J.~Khoury and
A.~Weltman, 
Phys.\ Rev.\ D \textbf{70}, 123518 (2004) 
[astro-ph/0408415]. 

\bibitem{vanstein} A.~I.~Vainshtein, 
Phys.\ Lett.\ \textbf{39B}, 393 (1972). 

\bibitem{Jwang} J.~Wang, L.~Hui and J.~Khoury, 
Phys.\ Rev.\ Lett.\ \textbf{109}, 241301 (2012) 
[arXiv:1208.4612 [astro-ph.CO]]. 

\bibitem{KBamba} K.~Bamba, R.~Gannouji, M.~Kamijo, S.~Nojiri and M.~Sami, 
JCAP \textbf{1307}, 017 (2013) 
[arXiv:1211.2289 [hep-th]]. 

\bibitem{dRGT1} C.~de Rham and G.~Gabadadze, 
Phys.\ Rev.\ D \textbf{82}, 044020 (2010) 
[arXiv:1007.0443 [hep-th]]. 

\bibitem{dRGT2} C.~de Rham, G.~Gabadadze and A.~J.~Tolley, 
Phys.\ Rev.\ Lett.\ \textbf{106}, 231101 (2011) 
[arXiv:1011.1232 [hep-th]]. 

\bibitem{galileon} A.~Nicolis, R.~Rattazzi and E.~Trincherini, 
Phys.\ Rev.\ D \textbf{79}, 064036 (2009) 
[arXiv:0811.2197 [hep-th]]. 

\bibitem{EXTENDED} S. Capozziello, R. D'Agostino, O. Luongo, (2019)
[arXiv:1904.01427 [gr-qc]].

\bibitem{CAPOZZI} S. Capozziello, S. Nojiri, S. D. Odintsov, A. Troisi,
Phys. Lett. B\textbf{\ 639}, 135 (2006) [arXiv:astro-ph/0604431v3].


\bibitem{KHOURY2016} L.~Berezhiani, J.~Khoury and J.~Wang, 
Phys.\ Rev.\ D \textbf{95}, no. 12, 123530 (2017) 
[arXiv:1612.00453 [hep-th]]. 

\bibitem{Farooq:2013hq} O.~Farooq and B.~Ratra, 
Astrophys.\ J.\ \textbf{766}, L7 (2013) [arXiv:1301.5243 [astro-ph.CO]]. And
the references their in.

\bibitem{planck} P.~A.~R.~Ade \textit{et al.} [Planck Collaboration], A \&
A, 571, A16 (2014) 
arXiv:1303.5076 [astro-ph.CO].

\bibitem{Suzuki:2011hu} N.~Suzuki, D.~Rubin, C.~Lidman, G.~Aldering,
R.~Amanullah, K.~Barbary, L.~F.~Barrientos and J.~Botyanszki \textit{et al.}
, 
Astrophys.\ J.\ \textbf{746}, 85 (2012) [arXiv:1105.3470 [astro-ph.CO]]. 

\bibitem{Lazkoz05} R. Lazkoz, S. Nesseris, L. Perivolaropoulos, J. Cosmol.
Astropart. Phys., 0511, 010 (2005)



\bibitem{Blake:2011en} C.~Blake, E.~Kazin, F.~Beutler, T.~Davis,
D.~Parkinson, S.~Brough, M.~Colless and C.~Contreras \textit{et al.}, 
Mon.\ Not.\ Roy.\ Astron.\ Soc.\ \textbf{418}, 1707 (2011) [arXiv:1108.2635
[astro-ph.CO]]. 


\bibitem{Percival:2009xn} W.~J.~Percival \textit{et al.} [SDSS
Collaboration], 
Mon.\ Not.\ Roy.\ Astron.\ Soc.\ \textbf{401}, 2148 (2010) [arXiv:0907.1660
[astro-ph.CO]]. 


\bibitem{Beutler:2011hx} F.~Beutler, C.~Blake, M.~Colless, D.~H.~Jones,
L.~Staveley-Smith, L.~Campbell, Q.~Parker and W.~Saunders \textit{et al.}, 
Mon.\ Not.\ Roy.\ Astron.\ Soc.\ \textbf{416}, 3017 (2011) [arXiv:1106.3366
[astro-ph.CO]]. 


\bibitem{Jarosik:2010iu} N.~Jarosik, C.~L.~Bennett, J.~Dunkley, B.~Gold,
M.~R.~Greason, M.~Halpern, R.~S.~Hill and G.~Hinshaw \textit{et al.}, 
Astrophys.\ J.\ Suppl.\ \textbf{192}, 14 (2011) [arXiv:1001.4744
[astro-ph.CO]]. 


\bibitem{Eisenstein:2005su} D.~J.~Eisenstein \textit{et al.} [SDSS
Collaboration], 
Astrophys.\ J.\ \textbf{633}, 560 (2005) [astro-ph/0501171]. 


\bibitem{Giostri:2012ek} R.~Giostri, M.~V.~d.~Santos, I.~Waga,
R.~R.~R.~Reis, M.~O.~Calvao and B.~L.~Lago, 
JCAP \textbf{1203}, 027 (2012) [arXiv:1203.3213 [astro-ph.CO]]. 

\bibitem{SAMI-NOJ} K. Bamba, Md. Wali Hossain, R. Myrzakulov, S. Nojiri, M.
Sami, Phys. Rev. D \textbf{89}, 083518 (2014) [arXiv:1309.6413 [hep-th]].
\end{thebibliography}
\end{document}